\documentclass{aa}
\usepackage{graphicx}
\usepackage{txfonts}
\usepackage{natbib,graphicx,ulem}
\begin{document}
   \title{Kinematics and Stellar Populations
          of Low-Luminosity Early-Type Galaxies in the Abell~496 Cluster 
    \thanks{Based on observations obtained with ESO VLT (program 074.A-0533)
and MegaPrime/MegaCam, a joint  project of Canada France Hawaii Telescope
(CFHT) and CEA/DAPNIA, at the
Canada-France-Hawaii Telescope (program 03BF12), which is operated by the
National Research Council (NRC) of Canada, the Institut National des
Sciences de l'Univers of the Centre National de la Recherche Scientifique
(CNRS) of France, and the University of Hawaii.}
\thanks{All the spectral and imaging data used in this paper
are available through the VO-Paris Data Centre:
http://voplus.obspm.fr/DataCollections/Abell0496/}
}

   \author{Igor V. Chilingarian\inst{1,2}
          \and
          Veronique Cayatte\inst{3}
         \and
          Florence Durret\inst{4}
           \and
          Christophe Adami\inst{5}
          \and
          Chantal Balkowski\inst{6}
          \and
          Laurent Chemin\inst{6}
          \and
          Tatiana F. Lagan\'a\inst{7}
          \and
          Philippe Prugniel\inst{8,6}
          }

   \offprints{Igor Chilingarian \email{igor.chilingarian@obspm.fr}}

   \institute{Observatoire de Paris-Meudon, LERMA, UMR~8112, 61 Av. de l'Observatoire, Paris, 75014, France
         \and
              Sternberg Astronomical Institute, Moscow State University, 13 Universitetski prospect, 119992, Moscow, Russia
         \and
             Observatoire de Paris-Meudon, LUTH, UMR~8102, 5 pl. Jules Janssen, Meudon, 92195, France
         \and
             Institut d'Astrophysique de Paris, CNRS, UMR~7095, Universit\'e Pierre et Marie Curie, 98bis Bd Arago, 75014 Paris, France
         \and
             Laboratoire d'Astrophysique de Marseille, UMR~6110, Traverse du Siphon, 13012 Marseille, France
         \and
             Observatoire de Paris-Meudon, GEPI, UMR~8111, 5 pl. Jules Janssen, Meudon, 92195, France
         \and
            Instituto de Astronomia, Geof\'{\i}sica e C. Atmosf./USP, R. do Mat\~ao 1226, 05508-090 S\~ao Paulo/SP, Brazil 
         \and
             Universit\'e de Lyon, Lyon, F-69000, France ; Universit\'e Lyon~1,
             Villeurbanne, F-69622, France; Centre de Recherche Astronomique de
             Lyon, Observatoire de Lyon, 9 av. Charles Andr\'e, St. Genis Laval, F-69561, France ; CNRS, UMR 5574 ;
             Ecole Normale Sup\'erieure de Lyon, Lyon, France\\
             }

   \date{Received 15/Feb/2008; accepted 12/Mar/2008; in original form 20/Sep/2007}

   \authorrunning{Chilingarian et al.}
   \titlerunning{Abell 496: Kinematics and Stellar Populations}

\abstract
{The morphology and stellar populations of low-luminosity early-type
galaxies in clusters have until now been limited to a few relatively
nearby clusters such as Virgo or Fornax. Scenarii for the formation and
evolution of dwarf galaxies in clusters are therefore not well
constrained.}
{We investigate here the morphology and stellar populations of
low-luminosity galaxies in the relaxed richness class 1 cluster Abell~496 ($z=0.0330$).}
{Deep multiband imaging obtained with the CFHT Megacam 
allowed us to select a sample of faint galaxies, defined
here as objects with magnitudes $18<r'<22$~mag within a 1.2~arcsec fibre 
($-18.8 < M_B < -15.1$~mag). We observed 118 galaxies spectroscopically with the
ESO VLT FLAMES/Giraffe spectrograph with a resolving power
$R=6300$. We present structural analysis and colour maps for the 48
galaxies belonging to the cluster. We fit the spectra of 46 objects with
PEGASE.HR synthetic spectra to estimate the
ages, metallicities, and velocity dispersions. We estimated possible
biases by similarly analysing spectra of $\sim$1200 early-type galaxies
from the Sloan Digital Sky Survey Data Release 6 (SDSS DR6). We
computed values of
$\alpha$/Fe abundance ratios from the measurements of
Lick indices. We briefly discuss effects of the fixed aperture size
on the measurements.}
{For the first time, high-precision estimates of stellar population
properties have been obtained for a large sample of faint galaxies in
a cluster, allowing for the extension of 
relations between stellar populations and
internal kinematics to the low-velocity dispersion regime.
We have revealed a peculiar population of elliptical 
galaxies in the core of the cluster, resembling massive early-type
galaxies by their stellar population properties and velocity dispersions,
but having luminosities of about 2~magnitudes fainter.}
{ External mechanisms of gas removal
(ram pressure stripping and gravitational harassment) are more 
likely to have occurred than internal mechanisms such as supernova-driven winds.
The violent tidal stripping of intermediate-luminosity, early-type galaxies
in the cluster core can explain the properties of the peculiar elliptical 
galaxies surrounding the cD galaxy.}

   \keywords{evolution of galaxies; galaxies: clusters: individual
(Abell 496); kinematics; stellar populations; spectroscopic survey }
\maketitle
%

\section{Introduction}

Understanding the formation and evolution of galaxies is one of the most
challenging tasks in modern astrophysics, and substantial progress has
been achieved in characterising the evolutionary pattern of early-type
galaxies.  Massive ellipticals, which are found principally in clusters,
are known to be already present at z $\simeq$ 1 (Ziegler 2000, and
references therein).  Recent large surveys give strong support to: (1)
the downsizing star formation concept (the star formation activity is
seen to progress with time, from high mass galaxies to smaller ones (as
first suggested by Matteucci 1994) from variations of the [Mg/Fe] ratio
with galaxy luminosity in ellipticals, then by Cowie et al. 1996); and
(2) the top-down formation where mass assembly occurs at lower redshifts
for lower galaxy masses (Bundy et al. 2006; Cimatti et
al. 2006). However, if a consensus exists for star formation timescales
and chemical evolution, it is not clear even for massive ellipticals if
the mass assembly time corresponds to the star formation episode
duration (Bell et al. 2006; Pozzetti et al.  2007; Scarlata et
al. 2007).  In addition, multiple mergers of smaller galaxies are not
the only formation mechanism, also leaving room for a rapid collapse of
gas in the remote past. Moreover, galaxy evolution in clusters is
expected to be different from that in lower density environments. The
local morphology-density relation revealed by Dressler (1980), with a
very high fraction of early-type galaxies found in cluster cores, has
been extended to a larger range of galaxy density from the SDSS (Goto et
al. 2003) and to higher redshift (Capak et al. 2007). Both the latter
study and the disc-fading model discussion for S0 galaxy formation by
Christlein \& Zabludoff (2004) indicate that the morphology and star
formation could be affected by different processes: the increase in
early-type fraction is mostly driven by galaxy interactions and
harassment (Moore et al. 1998) or by tidal effects induced by the
crossing of the cluster potential well (including dynamical friction);
the reduction or suppression of the star formation is caused by ram
pressure gas stripping (Gunn \& Gott 1972, Abadi et al. 1999) or by
strangulation, which prevents further gas accretion by cutting off the
outer neutral gas reservoir (Larson et al. 1980). In conclusion, the
bulges of lenticulars in clusters are suspected to be the results of
tidal interactions.

Such scenarii of morphological transformations of infalling galaxies are
also invoked for the transformation of dwarf irregulars or faint
late-type spirals into early-type dwarf galaxies constituting the most
numerous class of galaxies in nearby clusters. However, until now both
the harassment model and the classical wind model in a virialized
protogalaxy fail to reproduce the observed positions of dwarf
ellipticals in the fundamental plane (Djorgovski \& Davis, 1987) which
links the internal kinematics and the structural properties (De Rijcke
et al.  2005). Geha et al. (2003) and Van Zee et al. (2004) faced
similar difficulties trying to explain the intermediate ages and
slightly subsolar to solar metallicities found for dwarf ellipticals
(dEs) from the analysis of Lick indices: no evolutionary scenario could
be ruled out or confirmed. To complicate the situation, Lisker et
al. (2007) have shown that early-type dwarfs including ellipticals and
dwarf lenticulars (dS0s) do not form a homogeneous class of galaxy.  At least five
subclasses with different morphological and clustering properties are
found in the Virgo cluster: the dE(di)s displaying disc-like features
(Lisker et al. 2006a); the dE(bc)s showing a blue centre with recent or
ongoing star formation (Lisker et al. 2006b); the bright nucleated
dwarfs dE(N)s form an unrelaxed population of disc-shaped dwarfs, which
is suspected to be the result of transformations of infalling
progenitors; faint nucleated dwarfs; and all non-nucleated dEs form a
relaxed population of classical spheroidal objects that formed or
arrived in the cluster a long time ago.

In order to clarify the evolutionary path of intermediate and low-mass,
early-type galaxies in clusters we have studied the photometric,
stellar population and kinematical properties of a sample of dwarf
galaxies in the nearby cluster \object{Abell~496}. Our goal is to point out
objects that are clearly the results of the different proposed
scenarii to define which properties are discriminant for the formation
and evolution of faint early-type galaxies. It is more promising to identify
and select the most probable candidates rather than performing
statistical studies on the whole class of intermediate and low-mass
early-type objects.  A first step has been the discovery of a new
compact elliptical which is the result of tidal stripping by the
central cluster Dominant (cD) galaxy of an intermediate mass lenticular 
(Chilingarian et al. 2007c).

\object{Abell~496} is a richness class 1 cluster (Abell 1958) of cD type (Struble
\& Rood 1987) at a heliocentric velocity of 9885~km~s$^{-1}$
($z=0.0330$, Durret et al. 2000 and references therein), or
9707~km~s$^{-1}$ ($z_{corr}=0.0324$) after correcting for infall of the
Local Group toward Virgo. For a Hubble constant H$_0$=73~km~s$^{-1}$
Mpc$^{-1}$, the distance modulus is 35.70 ($d=133$~Mpc) and the
corresponding scale is 0.627~kpc~arcsec$^{-1}$.  \object{Abell~496} is a cluster
with several hundred measured galaxy redshifts (Durret et al. 1999). The
analysis of the distribution of 466 redshifts in the direction of this
cluster has revealed the existence of several structures along the line
of sight; however, the redshift distribution of the 274 galaxies found
to belong to the cluster itself implied that \object{Abell~496} has a regular
morphology and a well relaxed structure (Durret et al. 2000). This is
confirmed by X-ray data: the X-ray map obtained from XMM-Newton
observations is indeed quite regular, contrary to most clusters where
even if the X-ray emissivity map appears regular, the temperature map of
the hot gas does not (e.g., Durret et al. 2005).

We will present here results for 48 low-luminosity galaxies in
\object{Abell~496}.  The observations and data reduction are described in
Sect.~2. We give the photometric properties of the \object{Abell~496} sample in
Sect.~3 and a description of the spectral fitting applied to recover the
stellar population and kinematical properties in Sect.~4. We present
results from the photometric and spectroscopic analyses in
Sect.~5, and discuss our conclusions in Sect.~6.

\section{Observations and data reduction}
\subsection{Imaging observations and sample selection}

We obtained images with the Canada France Hawaii Telescope with the
Megacam camera in the fall of 2003 (program 03BF12,
P.I. V.~Cayatte). Megacam covers a field of 1$^\circ \times 1^\circ$ on
the sky, with a pixel size of 0.187$\times$0.187~arcsec$^2$. We obtained
deep images in the $u^*$, $g'$, $r'$, and $i'$ filters.  We reduced
these images in the usual way (bias and flat field corrections,
photometric, and astrometric calibrations) by the staff of the Terapix
data center at Institut d'Astrophysique de Paris (IAP), France. The
SExtractor software was run on the $r'$ image (the image with the best
seeing) to detect objects and measure their positions and magnitudes. In
particular, we measured magnitudes within a 1.2~arcsec diameter, 
to prepare FLAMES/Giraffe observations (see below).  Details on
the data reduction of these images can be found in Bou\'e et al. (2008).

We then discarded stars based on a diagram of aperture minus total
magnitude versus total magnitude for $r'<$21. Above this magnitude, we
kept all objects in our galaxy sample. A photometric redshift code was
kindly applied by O.~Ilbert to our catalogue in an attempt to eliminate
background galaxies. Finally, the galaxies observed with FLAMES/Giraffe
were taken from this imaging catalogue, with a magnitude within a
diameter of 1.2~arcsec in the $r$ band ($r'_{1.2}$) in the [18-22]
interval.

\begin{figure*}
\includegraphics[width=18cm]{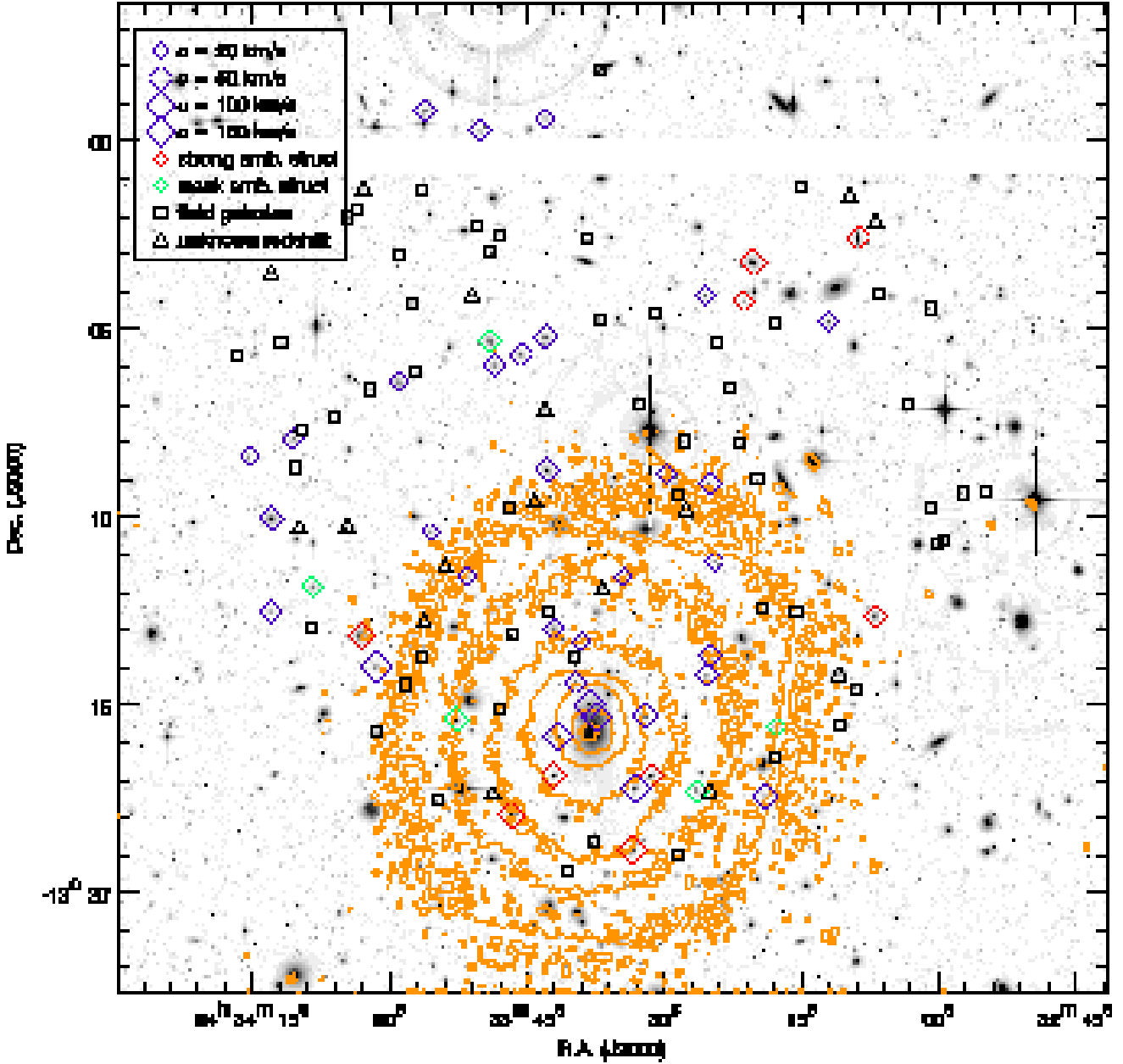}
\caption{A fragment of the CFHT Megacam $r$'-band image, showing the
central part of \object{Abell~496}. Galaxies from the spectroscopic sample are
displayed with different symbols: (1) confirmed cluster members as diamonds
with the colour indicating the presence of embedded structures: red, green,
and blue for no, weak, and strong embedded structures respectively; (2)
confirmed field galaxies as squares; (3) galaxies with unknown redshifts as
triangles. The X-ray map obtained with XMM-Newton is shown as contours.
\label{figa496cfht}}
\end{figure*}

We took the observed galaxies from the catalogue described above, with
the following priorities: top priority, objects with
$18.0<r'_{1.2}<20.75$; middle priority, objects with
$20.75<r'_{1.2}<21.5$; and low priority, objects with
$21.5<r'_{1.2}<22$. We thus obtained 118 galaxy spectra (some fibers had
to be used for guide stars and sky spectra).  A fragment of the
$r$'-band image with the galaxies observed spectroscopically with
FLAMES/Giraffe is shown in Fig.~\ref{figa496cfht}.

\subsection{Spectroscopic observations and reduction}
We obtained spectra with the ESO Very Large Telescope using the FLAMES/Giraffe
instrument in the L543.1 configuration on the two nights of 8-10/12/2004. The
FLAMES/Giraffe field of view is 20~arcmin in diameter, with a total number of
fibres of 130; each fiber has a circular aperture on the sky of 1.2~arcsec in
diameter. We used the 600 lines/mm grating in the LR4 setup, giving a
resolving power of about $R=6300$ in the wavelength range 5010--5831~\AA.

We obtained four exposures on the first night for 2700, 3300, 2351, and
1699 seconds. The second night four other exposures have been acquired
with the same positioner configuration file and effective durations of
2700, 3300$\times$2 and 4200 seconds. During the day, exposures of bias,
flat fields, and arc line lamps for the wavelength calibration have been
done in the same setup and with the two separate sets of MEDUSA
fibers. The description of the FLAMES/Giraffe instrument can be found in
Pasquini et al. (2002). We extracted and calibrated the spectra were
using the Python version of BLDRS - Baseline Data Reduction Software
(girbldrs-1.12) available from http://girbldrs.sourceforge.net and with
functions and recipes described in the BLDRS Software Reference Manual,
Doc. No. VLT-SPE-OGL-13730-0040 (Issue 1.12, 20/9/2004). We did the
reduction with the Image Reduction and Analysis Facility (IRAF). The
processing includes bias subtraction; diffuse light estimation;
removal, localisation, and extraction of the spectra; correction for
fiber transmission variations; wavelength calibration; division by the
continuum lamp spectrum; and sky subtraction. The resulting spectra are
not calibrated in absolute fluxes. We use the flux uncertainties
provided by the BLDRS for the data analysis.

We combined individual 1D spectra with the RSI IDL software and
we measured redshifts with the rvsao.xcsao package in IRAF, using
various stellar templates. Redshifts were also measured through the
stellar population synthesis fit described in this work, and both
values agreed within their uncertainties.

We had a very good result for our star-galaxy separation since we did
not observe any stars spectroscopically. On the other hand, the
rejection of background objects based on photometric redshifts was not
very efficient, since only 48 out of 102 galaxies with measurable
redshifts actually belong to the cluster. Of those 48, 46 have
sufficient signal-to-noise ratios to analyse their kinematic and stellar
populations. For the two remaining objects we present only the
photometric analysis.

Absolute magnitudes were computed using the distance modulus mentioned
above. All magnitudes considered throughout the paper are corrected for
intergalactic extinction according to Schlegel et al. (1998).  We
corrected for cosmological dimming, and applied the K-correction and
conversion into the $B$-band (if needed) assuming an elliptical galaxy
SED and transformations from Fukugita et al. (1995).

\section{Spectral fitting}
To deduce kinematical and stellar population parameters, we have used
direct fitting of the PEGASE.HR (Le Borgne et al. 2004) synthetic
spectra to observed data in pixel space (van der Marel \& Franx 1993,
Cappellari \& Emsellem 2004). At the same time, the minimisation
procedure returns the parameters of the population
model (age and metallicity of Simple Stellar Populations, hereafter
SSPs) and of the internal kinematics (Gaussian LOSVD).

Details of the method are given in Chilingarian et al. (2005, 2007d) and
Prugniel et al. (2005), and its stability and biases are described in
Chilingarian et al. (2007a).  In the present paper we fit the observations
with single SSPs computed with the Salpeter (1955) IMF. This gives us the
SSP-equivalent stellar population parameters that we will refer to
throughout the text. We use a 15th order multiplicative polynomial
continuum in the fitting procedure and no additive continuum, as discussed
in Appendix B, together with the effects of non-solar [Mg/Fe] abundance 
ratios on the stellar population parameters.

To obtain reliable and precise uncertainties of the stellar population
parameters, the computations are done in the rotated coordinate system
defined as $\eta = (3 Z + 2 \log_{10} t) / \sqrt{13}; \theta = (-2 Z + 3
\log_{10} t) / \sqrt{13}$, where the $\eta$ axis is parallel to the
direction of the age--metallicity degeneracy for intermediate-age and
old stellar populations noticed by Worthey (1994).

Since the spectral resolution of FLAMES/Giraffe in the MEDUSA mode is
rather high ($R=6300$), the PEGASE.HR models, based on the
high-resolution ($R=10000$) ELODIE.3 empirical stellar library (Prugniel
\& Soubiran 2001, 2004), remain the only choice if one tries to: (1)
avoid a degradation of the spectral resolution of the observed spectra;
and (2) use an empirical stellar library. In order to acquire unbiased
estimates of the velocity dispersions, one needs to take into account
variations of the spectrograph line-spread-function (LSF) and to broaden
the template spectra according to the LSF shape (strictly speaking,
according to the difference between the LSF of the spectrograph used to
obtain the spectra being analysed and that of the stellar library used
for spectral synthesis purposes). To achieve this, we fit the twilight
spectra obtained with FLAMES/Giraffe in the same setup as the \object{Abell~496}
galaxies with solar spectra available in the ELODIE.3 library, which
obviously have exactly the same intrinsic LSF as the stars used for the
spectral synthesis. The instrumental response of FLAMES/Giraffe appeared
to be very stable across the fibers. The instrumental width
($\sigma_{inst}$) changes smoothly from 19~km~s$^{-1}$ at 5000~\AA\ to
15~km~s$^{-1}$ at 5800~\AA, H3 remains stable at about $-0.01$, and H4
at about $-0.07$ (see the definition of Gauss-Hermite parametrization in
van der Marel \& Franx, 1993). Slightly negative values of H4 are
trivially explained by the fiber sizes (1.2~arcsec) which are larger
than a normal spectrograph slit width (diffraction limit of the
collimator) resulting in a $\Pi$-shaped LSF.  A high spectral resolution
of FLAMES/Giraffe allows us to measure velocity dispersions as low as
10~km~s$^{-1}$ for data having a signal-to-noise ratio of about 10 per
pixel.

Since our technique for extraction of stellar population and kinematical
parameters is based on a non-linear, least-square fitting on many
parameters, there is always a chance that the minimisation procedure
does not reach the absolute minimum in the $\chi^2$ space. In addition
there are several degeneracies between the parameters, the most
important being the (1) age-metallicity, and (2) metallicity-velocity
dispersion parameters. For this reason, the shape of the minimum in the
parameter space becomes strongly extended along the lines corresponding
to those degeneracies, sometimes exhibiting several local minima, where
the minimisation algorithm can be trapped.
To check how critical the problem is in our case, we built $\chi^2$ maps.

We proceeded as in Chilingarian et al. (2007a) --
Appendix A -- fitting only the kinematics and a multiplicative
polynomial continuum for a set of fixed values of ages ($t$) and
metallicities ($Z$) of the templates; thus for every pair of values
($t, Z$) the best fitting kinematical parameters are
obtained. Finally, we obtained a map of minimal $\chi^2$ values in the
age--metallicity space for each spectrum. These maps allow us to detect
possible systemic errors on the stellar population parameters derived
from the non-linear fitting.

For the 46 spectra the solutions of the fitting coincide with the minima
seen on the maps, therefore, our minimisation strategy can be considered
as reliable (see figures in Appendix A).  The reduced $\chi^2$ values
for the spectra ($\approx 0.45$), where no template mismatch due to
non-solar [Mg/Fe] ratios is seen, suggest that the flux uncertainties
provided by the BLDRS are overestimated by $\sim$50 percent.

\section{Results}

\subsection{Photometric and morphological properties}
We provide integrated photometric parameters and colour maps
for the 48 galaxies identified as definitive members of the cluster. 
Surface photometry, profile decomposition and
fundamental relations will be discussed in a forthcoming paper. As
previously discussed, the galaxies selected inside the virial radius of a
relaxed cluster and in its redshift range have a large probability of being
early-type galaxies, i.e., ellipticals or lenticulars.

From the Megacam images, we perform a simplified morphological
classification by visual inspection, determining if the galaxy is an
elliptical, S0, Sa, or late-type spiral.  Morphological types are always
evaluated with a scatter of about one type given the uncertainty on this
measurement, so, in some cases, we kept two possible types. This
classification is reported in Table~1 of Appendix~A. We have found only
two galaxies to be of a later type than Sa and one has been classified
as SBa. The 45 other galaxies can be considered as real, early-type
galaxies.  Even if our sample is not completely representative of the
whole population of the cluster because some observational constraints
(avoidance of bright objects and some limitations due to the
FLAMES/Giraffe positioner) have been set, the morphological segregation
is observed as expected in a cluster core.

Since the range in absolute $B$ magnitude [$-18.8,-15.0$] covers the
limit between intermediate luminosity and dwarf galaxies, we split the
sample into three different subsets: the brightest representatives
corresponding to the intermediate luminosity galaxies of Bender et
al. (1992) with $M_{B}<-18.0$ (among our full sample ten galaxies belong
to this subset), a transition subsample with $M_{B}$ between $-18.0$ and
$-17.5$ mag (the classification is given as dS0/S0 or dE/E depending
whether the galaxy is found to be lenticular or elliptical; this subset
contains eight galaxies), and finally, the real, early-type dwarf set
with $M_{B}>-17.5$ (28 objects, i.e., 57 percent of the whole
sample). Among all three subsets, the galaxies are at different
projected distances from the cD (taken as the centre of the cluster).
The morphological classification seems pertinent for the brightest
objects, but we will not discuss the separation between dS0 and dE, as
suggested by Lisker et al. (2007) who assigned a common abbreviation
``dE'' to this rather heterogeneous class of faint galaxies.  They
concluded their study by evidencing various subclasses of dEs, and we
will examine our sample in such a context.

We have applied an elliptically-smoothed unsharp masking technique
(e.g. Lisker et al. 2006a) to the CFHT/Megacam images of our galaxies to
search for embedded structures. Using different smoothing radii (semi
major-axes of ellipses) from 0.3 to 4 arcsec, we classified all the
objects into three categories: no, weak, and strong embedded structures.
They included bar, disc, spiral arms, and ring types (see
Tab~\ref{tabmorph}). Nine and five objects were found to have strong and
weak embedded structures, respectively. We stress that only one galaxy
with $M_{B}>-17.5$ shows strong embedded structures and three other
faint $-17.5<M_{B}<-16.5$ objects have weak ones; on the other hand,
brighter objects often exhibit strong and complex structures, not
observed in fainter early-type dwarfs.

Among the nine objects with strong embedded structures two sets of
three galaxies are found relatively near each other in projection but
with very different radial velocities. The first set
(ACO496J043308.85-130235.6 (G-02), ACO496J043320.35-130314.9 (G-06)
and ACO496J043321.37-130416.6 (G-07)) is located at the northern limit
of the X-ray halo (see Fig.~12 of Tanaka et al. 2006); the second
set (ACO496J043326.49-131717.8 (G-13), ACO496J043331.48-131654.6
(G-15) and ACO496J043333.53-131852.6 (G-18)) is in the south near the
cluster centre, in a region where the X-ray surface brightness is seen
in excess compared to its azimuthally averaged value and corresponding
to a cold front. However, we interpret the set of three galaxies
projected on the cluster core as three objects only seen close in
projection (see notes for each galaxy in Appendix~A). In both cases,
the existence of a real group of galaxies has to be tested.

Concerning the galaxies with weak embedded structures, one of them is
the brightest object found among a group of four galaxies in our
spectroscopic sample, which are all in the same region of the sky about
400~kpc north of the cD. As discussed in the notes of Appendix~A,
ACO496J043348.59-130558.3 (G-32), ACO496J043343.04-130514.1 (G-28),
ACO496J043345.67-130542.2 (G-30) and ACO496J043349.08-130520.5 (G-33) probably
belong to a real group since their radial velocities are close to each
other.

Finally, eight dwarf galaxies with $M_{B}>-18.0$~mag exhibit spiral arms,
bars, rings, or edge-on discs similar to the structural properties
defined by Lisker et al. (2006a) in the dwarf population of the Virgo
cluster, corresponding to 24 percent for a total number of
34 dEs in the spectroscopic sample of \object{Abell~496}; this value is larger
than the percentage given by Lisker et al. (2007) for their dE(di)
subclass, but it can be explained by a bias toward bright galaxies
in our sample.  Another subclass pointed out by Lisker et al. is that of
dEs with blue centres which exhibit recent or ongoing central star
formation. Three objects are seen on the colour maps with such a
property; all of them are rather flattened galaxies compared to the
other dwarfs and could perfectly correspond to the flattened spheroid
invoked by Lisker et al. (2006b) to explain their subclass of dEs with
blue cores.  A last subclass defined in the Virgo cluster is obtained
by separating the ``featureless'' dE class into nucleated and
non-nucleated galaxies. Since Abell 496 is more distant than Virgo,
the identification of a nucleus is more difficult to define with the
same precision in terms of flux and size. From unsharp
masks we can identify the galaxies where strong gradients are observed
in the central regions; nine dwarf galaxies with $M_{B}>-18.0$~mag present
bright compact components in their centres, one of them having a
bright blue core.  Compared to the Virgo dwarf study, the estimated
number of nucleated objects missed could be larger because we cannot be sure
that small size nuclei are not smoothed by the seeing effects.  In the following
subsection we will discuss the results of the spectroscopic data in
terms of age of the central stellar population; another
subclassification will be proposed to provide tests for
formation and evolution scenarii.

\begin{figure}
\includegraphics[width=8.8cm]{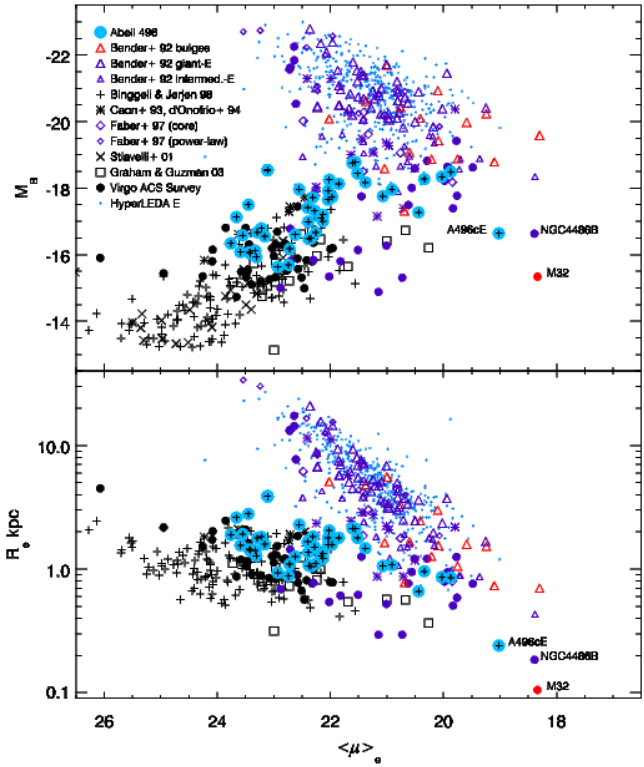}
\caption{Absolute B magnitude (top panel) and effective radius (bottom
panel, cf.  Kormendy relation) as a function of mean surface
brightness within the effective radius. Giant and
intermediate-luminosity ellipticals, power-law and core galaxies are
shown in blue, bulges of disc galaxies in red, dwarf ellipticals and
lenticulars in black. We keep the original morphological
classification (E/dE) for the data points coming from the Virgo
Cluster ACS Survey, and from Caon et al. (1993), and D'Onofrio et
al. (1994). Therefore, they appear both in blue and black. \object{Abell~496}
galaxies are shown as filled blue circles with crosses.
\label{figmuav}}
\end{figure}

Figure~\ref{figmuav} presents the updated versions of Figs.~9a,g
from Graham \& Guzm\'an (2003): absolute $B$ magnitude and effective radius
$R_e$ versus mean $B$ surface brightness within the 
effective radius $\langle\mu\rangle_e$, with
the \object{Abell~496} galaxies superimposed. The plot contains
literature data \textit{only} for elliptical galaxies and bulges of
lenticulars/spirals (where the bulge/disc decomposition has been made in the
original papers). All integrated measurements for spiral and lenticular
galaxies have been excluded. Data in computer-readable format for dE and E
galaxies from Binggeli \& Jerjen (1998), Caon et al. (1993), D'Onofrio et
al. (1994), Faber et al. (1997), Graham \& Guzm\'an (2003), Stiavelli et al.
(2001) and homogenization algorithms for these datasets have been kindly 
provided by A.~Graham. We also included photometric parameters of E and
dE/dS0 galaxies from the Virgo Cluster ACS Survey (Ferrarese et al.  2006);
photometric data on giant, intermediate elliptical galaxies and bulges of
spirals and lenticulars from Bender et al. (1992); photometric parameters of
the S\'ersic component of \object{M~32} (Graham 2002); and data for 430
elliptical galaxies from the HyperLeda\footnote{http://leda.univ-lyon1.fr/} 
database (Paturel et al. 2003), with radial velocities
below 10000~km~s$^{-1}$ and brighter than $M_B=-18.0$~mag. 

The well-known structural dichotomy between diffuse dwarf galaxies and
classical ellipticals and bulges is clearly visible on both plots in
Fig.~\ref{figmuav}. A sequence of classical elliptical galaxies starts with
the most luminous cluster galaxies at fainter surface brightnesses and ends 
with a few compact elliptical galaxies (M~32-like objects) at high surface
brightnesses. At the same time, diffuse dwarf elliptical galaxies form a
separate sequence. We notice that the counter-arguments againt this
interpretation of the structural diagrams exist (e.g. Graham \& Guzm\'an
2003). It is remarkable that over a large span of absolute
magnitudes ($-18 < M_B < -13$~mag) diffuse galaxies show no correlation
between effective surface brightness and effective radius: $r_{eff}$ always
remains between 0.6 and 2.0~kpc (see bottom panel). A number of objects, the
brightest representatives of our sample, lie in the transition region between
the two regimes (``normal'' and ``diffuse'' ellipticals).

\subsection{Kinematical properties}
\begin{figure}
\includegraphics[width=8.8cm]{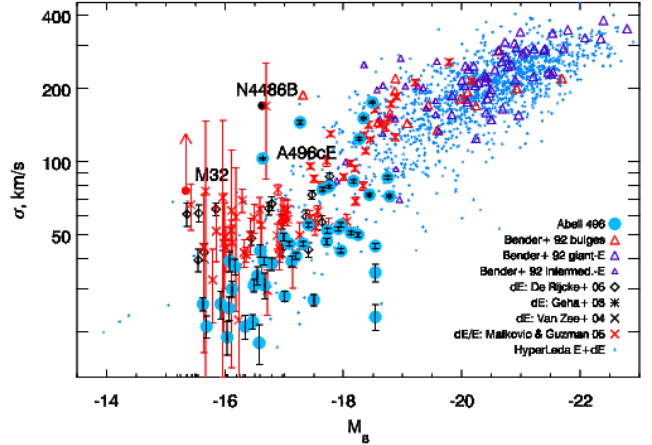}
\caption{Faber--Jackson relation for ellipticals and bulges of disc 
galaxies. \object{Abell~496} objects are shown as blue-filled circles.
Giant and intermediate-luminosity ellipticals are in blue, bulges
of disc galaxies in red, dwarf ellipticals and lenticulars in black. The
upper end of the arrow representing \object{M~32} corresponds to the HST
STIS measurements (Joseph et al. 2001), and the filled circle represents the
value obtained from earlier HST FOS data (van der Marel et al. 1998), which
was in agreement with more recent ground-based observations.
\label{figfjr}}
\end{figure}

The Faber-Jackson relation (Faber \& Jackson 1976), reflecting the
connection between the dynamical and stellar masses, is shown in
Fig.~\ref{figfjr}. We present a compilation of data for dwarf (Geha et
al. 2003; van Zee et al. 2004; De Rijcke et al. 2005; Matkovi\'c \&
Guzman, 2005), intermediate luminosity, and giant elliptical galaxies
and bulges of bright lenticulars (Bender et al.  1992).

Matkovi\'c \& Guzm\'an (2005) have analysed a sample of mostly dwarf and
low-luminosity
elliptical galaxies (${\rm M_B}<-18$) in the Coma cluster. In order to
have accurate B magnitudes for these galaxies, we retrieved the
photometric catalogue of galaxies in the direction of
Coma\footnote{http://cencosw.oamp.fr/} (Adami et al. 2006) and
cross-correlated it with the Matkovi\'c \& Guzm\'an catalogue. Values
for 66 matched objects have been used in Fig.~\ref{figfjr}.

Some objects of our sample located below the bulk of galaxies in the
luminosity range between $M_B=-17$ and $-19$~mag are mostly lenticular
galaxies with large discs, therefore, their velocity dispersions tend
to be lower (e.g., ACO496J043306.97-131238.8 (G-01)).

We observe a population of objects with higher velocity dispersions than
expected for their luminosities. Most of them are located
in the inner 80~kpc from the cluster centre and exhibit quite unusual
stellar populations. We will discuss their origin and evolution below.

We found a number of \object{Abell~496} objects, located systematically below
dE galaxies in the literature. Since most of the studies
(apart from Geha et al. 2003) were based on spectroscopy with
significantly lower spectral resolution than FLAMES/Giraffe, we cannot
exclude the possibility of systematic errors on velocity dispersions for
low-$\sigma$ objects in the literature due to a template mismatch
($\sigma$-metallicity degeneracy, see Chapter~1 in Chilingarian 2006).

\subsection{Stellar population properties}

Our fitting method is not sensitive to the presence or absence of the
H$\beta$ feature in the spectral range: though age estimations have
higher uncertainties when H$\beta$ is not included, they remain
unbiased (see Appendix~B for details).

To fit the spectra we use the PEGASE.HR synthetic populations built
using the ELODIE.3 library including only stars in the nearest solar
neighborhood.  These stars are known to have [Mg/Fe] correlated with
their [Fe/H] metallicities (see Wheeler et al. 1989 and references
therein). Thus fitting spectra of galaxies with non-solar [Mg/Fe] for
metallicities higher than $-1.0$~dex results in a template mismatch
that can bias our estimations of stellar population parameters. To
quantify possible biases and study the effect in detail we used
spectra of early-type galaxies from the Sloan Digital Sky Survey Data
Release 6 (Adelman-McCarthy et al. 2008). See Appendix~B for details.

To obtain [Mg/Fe] abundance ratios for the \object{Abell~496} galaxies and to
check whether our age and metallicity estimations are correct we used
stellar population models dealing with Lick indices of magnesium and
iron (Thomas et al. 2003).  In order to compute Lick indices we
degraded the spectral resolution to match that needed to compute Lick
indices (Worthey et al. 1994, Thomas et al. 2003) by convolving the
original spectra with a Gaussian having a width equal to the square
root of the differences between the squares of the Lick resolution 
($\sigma_{Lick}$), LSF ($\sigma_{inst}$) and velocity dispersion 
values found by the spectral fitting procedure ($\sigma_{g}$): 
$\sigma_{degr} = \sqrt{\sigma_{Lick}^2 - \sigma_{FL}^2 - \sigma_{g}^2}$.
One should also apply a velocity dispersion
correction for large galaxies, when the intrinsic broadening of the
spectral lines exceeds the Lick resolution. All the objects in our sample
have relatively low velocity dispersions, therefore these corrections
were not necessary. A similar approach to compute Lick indices has been used
by Kuntchner et al. (2006).
The spectral range of FLAMES/Giraffe in the setup we
used and the mean heliocentric redshift of the cluster $z=0.0330$ allow 
us to compute
the following Lick indices: Fe$_{5015}$, Mg$b$, Fe$_{5270}$,
Fe$_{5335}$, and Fe$_{5406}$. Uncertainties on the measurements were
computed according to Cardiel et al. (1998). 

There are three sources of systematic errors that can affect the
measurements of Lick indices: (1) data reduction issues resulting in
additive errors, such as problems with the diffuse light or sky subtraction;
(2) difference in the spectral resolution between given observations and the
Lick/IDS system; (3) uncertainties in the determination of the radial
velocities. 

The diffuse light subtraction for FLAMES/Giraffe is done with a high
accuracy thanks to the rather sparse packing of the MEDUSA fiber traces
on the CCD plane, therefore, it cannot result in significant errors.  We
observed the sky simultaneously with the galaxies in different parts of
the field of view, so it is possible to assess the quality of sky
subtraction by comparing individual sky spectra obtained in different
fibers. No systematic difference is observed, we therefore conclude that
the sky subtraction procedure is reliable.

The radial velocities and velocity dispersions of the \object{Abell~496} galaxies
are measured with high accuracy, due to the fact that the the
FLAMES/Giraffe LSF has almost infinite resolution in terms of square
differences, compared to the Lick/IDS system.

We also notice, that in the course of this study to analyze Lick indices
for both FLAMES/Giraffe and SDSS spectra (see Appendix~B), we use the
models by Thomas et al. (2003), rather than the models by Worthey et
al. (1994). Therefore, converting the computed values of Lick indices
into the Lick/IDS system through the observations of the Lick/IDS stars
is \textit{not necessary}. Moreover it was not possible to perform
this empirical conversion because Lick stars have not been observed.

The measurements of the Lick indices are presented in Appendix~A
(Table~\ref{tablick}).

There have been a number of recent studies (Gallazzi et al. 2006; Smith
et al. 2006; Yamada et al. 2006; Carretero et al. 2007) addressing
correlations between galaxy absorption line strengths and velocity
dispersions.  Most of them are based on observations
obtained with multi-object spectrographs with fixed spatial aperture
sizes (as in the present study). Although this observational
technique allows for the acquisition of spectra of many objects during a single
exposure time, it has some disadvantages when studying stellar
populations. Early-type galaxies (giants and dwarfs) and bulges of
spirals are known to have rather strong metallicity gradients
(see, e.g., S\'anchez-Bl\'azquez et al. 2006c) and often evolutionary decoupled
nuclei (e.g., Sil'chenko 2006; Chilingarian et al. 2007b; 2008b; Peletier et
al. 2007). Young and/or metal-rich stellar populations in nuclei
will not strongly affect the aperture measurements for galaxies
located 100~Mpc or further, because their spatial sizes are small and
contributions to the total fluxes within a 1.5~arcsec-wide aperture
are negligible, metallicity gradients may bias the results quite
strongly.

If members of a given cluster of galaxies are analyzed, the scale of the
gradient will depend on their effective radii, which are rather tightly
connected to luminosities. At the same time since luminosities of
early-type galaxies correlate with central velocity dispersions (Faber
\& Jackson 1976), for low-$\sigma$ objects the same aperture size in
average will contain a larger fraction of the galaxy, leading to an
underestimation of the central or effective metallicity. This may cause
a change in the slope of the Mg$b$-$\sigma$ relation, especially for
fainter objects (making it steeper). On the other hand, the velocity
dispersion also varies with radius; however, for low-luminosity and
dwarf galaxies, $\sigma$-profiles turn to be almost flat (Simien \&
Prugniel 2002; Geha et al. 2002; Geha et al. 2003; van Zee et
al. 2004). Therefore, while the aperture effect may not be very
important for giant galaxies (as mentioned in Gallazzi et al. 2006)
because the $\sigma$ gradient will compensate the metallicity gradient;
it may be important for dwarfs and intermediate luminosity objects.

When many clusters of galaxies at different redshifts are observed
with the same instrument (e.g., Gallazzi et al. 2006; Smith et al.
2006) the aperture effect will increase the spread of absorption line
strength measurements for a given velocity dispersion.

\begin{figure}
\includegraphics[width=8cm]{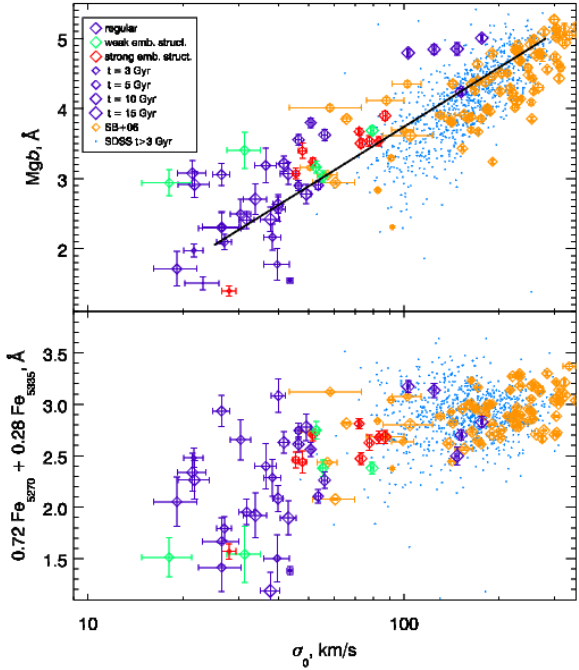}
\caption{Mg$b$ - $\sigma_0$ (top) and $\langle$Fe$\rangle$ - $\sigma_0$ (bottom)
relations.  Only points having $\Delta(\mbox{Mg}b)<0.5$~\AA\ are
shown. SDSS galaxies with ages older than 3~Gyr are shown as
light-blue dots. The black-solid line represents the correlation
between Mg$b$ and $\sigma_0$ found by Smith et al. (2006).
\label{figmgsig}}
\end{figure}

In Fig.~\ref{figmgsig}, we present the measurements of Mg$b$ and
$\langle$Fe$\rangle$ = 0.72~Fe$_{5270}$ + 0.28~Fe$_{5335}$ versus
velocity dispersion. We also put measurements for $\sim$700 early-type
galaxies older than 3~Gyr with redshifts $z<0.033$ observed in the SDSS
on the same plots. We have used values of Lick indices provided by the
SDSS. The solid line on the Mg$b$-$\sigma$ diagram corresponds to the
best-fitting relation from the National Optical Astronomical Observatory
(NOAO) Fundamental Plane Survey (NFPS, Smith et al. 2006). Measurements
for a sample of early-type galaxies from S\'anchez-Bl\'azquez et
al. (2006a,b,c), kindly provided by P. S\'anchez-Bl\'azquez in a
computer readable form, are shown in orange. There is relatively good
agreement between NFPS, SDSS, S\'anchez-Bl\'azquez et al. (2006a), and
the correlation we find for \object{Abell~496} galaxies. Our objects tend to be
slightly richer in magnesium, in contrast to the SDSS ones.
In our case, we have observed mostly galaxies near
the \object{Abell~496} cluster core, while in the case of NFPS, galaxies have been
observed in the peripheral parts of the clusters as well, and for SDSS
there was a number of group and field galaxies (since we did not apply
any environmental selection criteria). Smith et al. (2006; see also
Sil'chenko 2006, for an application to lenticular galaxies) 
demonstrated that objects in the cores of clusters tend to be above the
average line, and conversely, galaxies populating less dense environments
are less metal-rich.

We notice four high-$\sigma$ outliers from the \object{Abell~496} sample
significantly above the sequence of Smith et al. (2006) in the
Mg$b$-$\sigma$ plot. Those objects (ACO496J043333.17-131712.6 (G-17);
ACO496J043337.35-131520.2 (G-20); ACO496J043338.22-131500.7 (G-21);
ACO496J043341.69-131551.8 (G-24)) are all located in the innermost part of
\object{Abell~496} and have probably experienced strong tidal harassment (see
discussion).

\begin{figure}
\includegraphics[width=8.8cm]{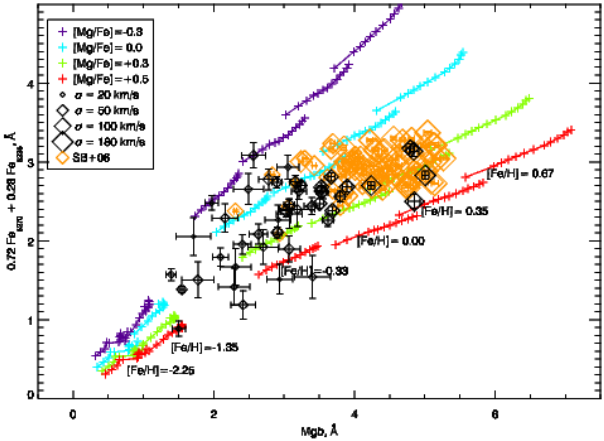}
\caption{ $\langle$Fe$\rangle$ vs Mg$b$ with sizes of symbols indicating the
values of central velocity dispersions. Models from Thomas et
al. (2003) for different values of [$\alpha$/Fe] enrichment are
overplotted. Only points having $\Delta(\mbox{Mg}b)<0.5$~\AA\ are
shown. 
\label{figmgfe}}
\end{figure}

The $\langle$Fe$\rangle$ versus Mg$b$ relation is shown in
Fig.~\ref{figmgfe}. The sizes of symbols reflect values of the velocity
dispersion. The data for early-type galaxies from S\'anchez-Bl\'azquez
et al. (2006a,b,c) are overplotted. The models of Thomas et al. (2003)
for $-0.3<$[Mg/Fe]$<+0.5$~dex are shown as crosses; each sequence
includes models for an indicated metallicity and ages from 3 to 15~Gyr
(bottom-left to top-right). Nearly all objects with low velocity
dispersions exhibit solar [Mg/Fe] ratios (the corresponding models are
shown in cyan). High-$\sigma$ objects have [Mg/Fe]$>$0.2~dex and, in
general, tend to have higher metallicity values.

In Appendix~B, we demonstrate that non-solar [Mg/Fe] ratios bias neither
age, nor metallicity estimations obtained by spectral fitting in the
wavelength range of FLAMES/Giraffe. Indeed, the measurements do suffer
from the well-known age-metallicity degeneracy (see, e.g., Worthey 1994)
expressed as $\Delta t / \Delta Z \approx 3/2$ or $\Delta (\log_{10}{t})
/ \Delta Z \approx 2/3$. In Appendix~A, we provide maps of $\chi^2$ in
the age-metallicity space for every object in our sample. The elongated
shape of the $\chi^2$ minima, corresponds exactly to the expected
degeneracy.

\begin{figure} 
\includegraphics[width=8cm]{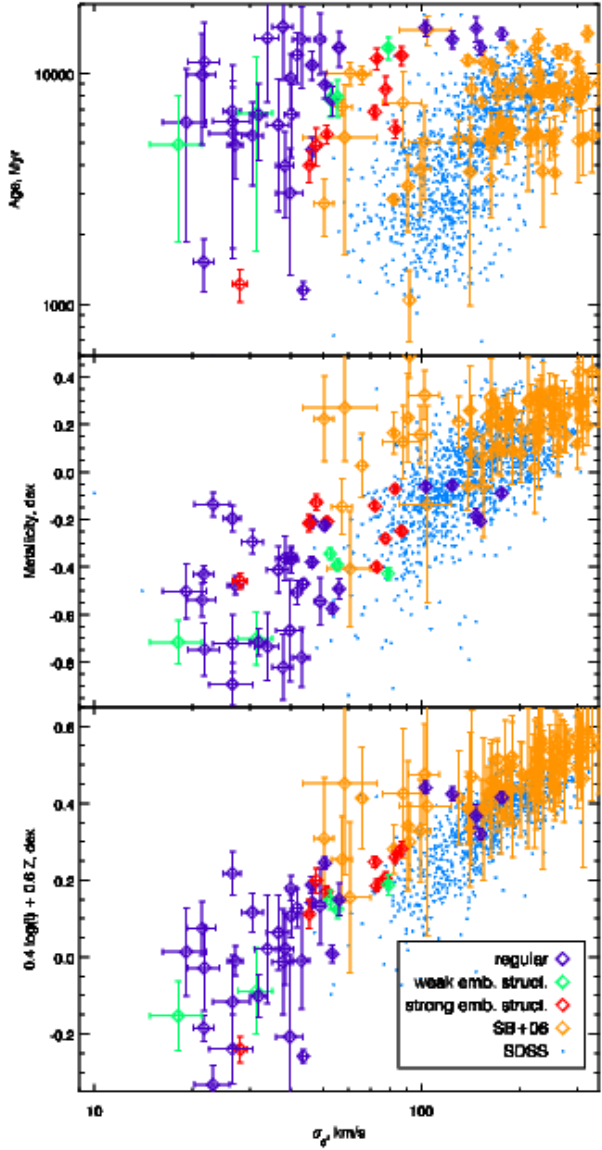}
\caption{Relations between central velocity dispersion and stellar
population parameters: SSP-equivalent age (top), metallicity (middle),
and a combination of these parameters free of the effect of the
age-metallicity degeneracy (bottom). The presence of embedded
structures in the galaxies is indicated. Values for early-type
galaxies from S\'anchez-Bl\'azquez et al.  (2006a,b) and measurements
for 848 SDSS galaxies are overplotted. Colours are the same as in
Fig.~\ref{figmgsig}.
\label{figtZsig}}
\end{figure}

In Fig.~\ref{figtZsig}, we compare the age $t$ and metallicity
estimates versus velocity dispersion for \object{Abell~496} galaxies and for
the samples of SDSS objects mentioned above and early-type galaxies
from S\'anchez-Bl\'azquez et al.  (2006a). We used the values
from S\'anchez-Bl\'azquez obtained by inverting the grid of the
H$\beta$ and $\langle$MgFe$\rangle$ Lick indices. The top panel presents $t$
versus $\sigma$. \object{Abell~496} galaxies are
systematically older than objects from the SDSS sample and the spread
of ages is quite large. However, we note (a) the absence of
young galaxies with high velocity dispersions; (b) the fact that small
galaxies (with low velocity dispersions) tend to be younger than large
ones;
and (c) the spread of age estimates is larger for low-mass objects. Two
explanations for the offset between the \object{Abell~496} galaxies and the two
other samples on the top panel are the environmental effects and the
sample selection. We have selected the SDSS galaxies with a minimal
signal-to-noise ratio, so dwarf galaxies have been nearly automatically
excluded. On the other hand, we confirm from our sample (which includes
the galaxies in the central region of a massive cluster) that
intermediate-luminosity objects tend to be older in dense environments,
since we find very few objects there younger than 10~Gyr having $\sigma
> 60$~km~s$^{-1}$.

Metallicity versus $\sigma$ (middle panel of Fig.~\ref{figtZsig})
exhibits a much stronger correlation than age vs $\sigma$. The \object{Abell~496}
galaxies appear to be almost on the low-$\sigma$ extension of the
sequence of the SDSS galaxies, although there are several faint
\object{Abell~496} dwarfs with rather high metallicities.  The bottom panel in
Fig.~\ref{figtZsig} shows a combination of age and metallicity giving an
edge-on view of the age-metallicity degeneracy: $0.4 \log_{10} t + 0.6
Z$ versus velocity dispersion. The correlation becomes much tighter than
$Z$-$\sigma$, and loci of \object{Abell~496} and SDSS galaxies follow the same
correlation. The spread of points (standard deviation) for a given value
of the velocity dispersion is less than 0.1~dex. A number of outliers in
the \object{Abell~496} sample are seen: two galaxies in the high-$\sigma$ area,
located in the very centre of the cluster (A496cE and A496g1 using the
terminology from Chilingarian et al. 2007c); and three dwarf galaxies
(ACO496J043324.61-131111.9 (G-08); ACO496J043339.07-131319.7 (G-22); and
ACO496J043355.55-131024.9 (G-37)). This quantity, $0.4 \log_{10} t + 0.6
Z$, is considered an indicator of the average strength of
absorption lines in the spectrum.

There are seven galaxies in our sample exhibiting relatively young
stellar populations ($\mbox{t} < 3$~Gyr): ACO496J043321.37-130416.6
(G-07); ACO496J043325.54-130408.0 (G-12); ACO496J043334.54-131137.1
(G-19); ACO496J043350.17-125945.4 (G-34); ACO496J043351.54-131135.5
(G-35); ACO496J043356.18-125913.1 (G-38); and ACO496J043415.37-130823.5
(G-46).  Three of them, ACO496J043321.37-130416.6 (G-07);
ACO496J043356.18-125913.1 (G-38); and ACO496J043415.37-130823.5 (G-46)
have narrow [OIII] emission lines in the spectra ($\sigma$ between 20
and 40 km~s$^{-1}$) and there is no evidence for the [NI]
($\lambda$=5199\AA) line in at least two of them, suggesting ongoing
star formation rather than shocked gas.  The first two, as well as
ACO496J043351.54-131135.5 (G-35), contain blue spatially unresolved
central regions clearly visible on colour maps. This can be considered
as an argument for the presence of young stars and/or star formation
only in the cores of the galaxies. All seven galaxies are located quite
far from the cluster centre (projected distances are 150 -- 200~kpc for
ACO496J043334.54-131137.1 (G-19) and ACO496J043351.54-131135.5 (G-35)
and $>450$~kpc for the other five).

A significant fraction of galaxies (13 of 46) exhibit red spatially
unresolved cores in the colour maps. Stellar populations of these
objects determined by spectral fitting are rather old and metal rich,
none of the spectra shows emission lines. In some cases
(ACO496J043320.35-130314.9 (G-06); ACO496J043331.48-131654.6 (G-15);
ACO496J043333.53-131852.6 (G-18); ACO496J043342.10-131653.7 (G-25);
ACO496J043346.71-131756.2 (G-31); and ACO496J043401.57-131359.7 (G-40))
the red cores represent the central parts of extended bar-like
structures, clearly visible on colour maps, which are redder than the
outer parts of their host galaxies.  However, in most of the other
red-core galaxies only the cores show peculiar colours, while the
discs/spheroids look more or less uniform on the maps. In the case of
ACO496J043306.97-131238.8 (G-01), an edge-on disky galaxy, dust
absorption is responsible for the redder colour of the central region.

\begin{figure}
\includegraphics[width=8cm]{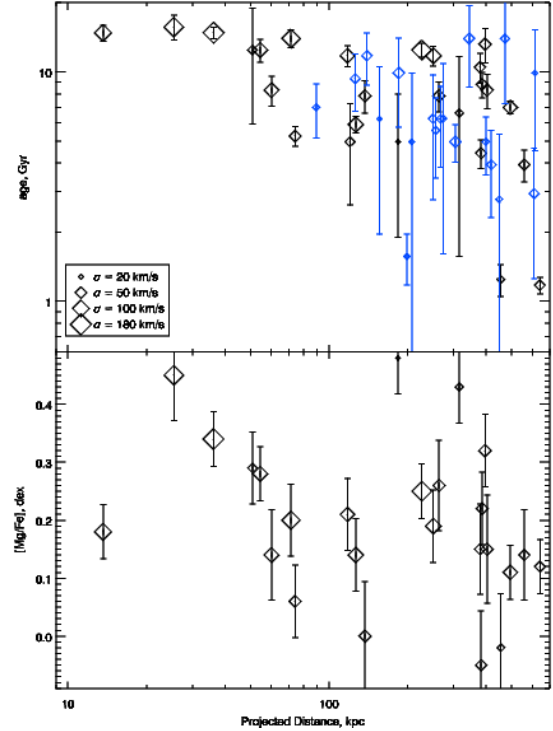}
\caption{Age and [Mg/Fe] abundance ratios of galaxies as a function of
projected distance to the cluster centre. Only galaxies having
uncertainties of [Mg/Fe] determinations better than 0.1~dex are shown in the 
bottom panel. The remaining galaxies are shown in blue in the top panel.
\label{figdproj}}
\end{figure}

Trying to relate the presence of embedded structures (discs, spiral
arms, bars) to the stellar population parameters, we stress that the
majority of the galaxies exhibiting embedded structures have
luminosity-weighted ages of between 4 and 10~Gyr (only
ACO496J043346.71-131756.2 (G-31) and ACO496J043403.19-131310.6 (G-41),
with bars and faint spiral arms, are about 14~Gyr old, but their
absolute blue magnitude is below $-18$~mag; ACO496J043321.37-130416.6
(G-07) is a young star-forming galaxy) and all of them have features
visible on colour maps. No objects with embedded structures are found in
the very inner part of the cluster ($d_{proj} < 60$~kpc) where the
luminosity-weighted ages of the five galaxies in our sample are above
12~Gyr. In Fig.~\ref{figdproj}, the ages and $\alpha$-enhancements of
galaxies at different projected distances from the \object{Abell~496} centre are
shown. In a transition area ($d_{proj}$ between 60 and 230~kpc) the age
of the youngest stellar populations decreases when the projected
distance increases, as it can be seen in Fig.~\ref{figdproj}.  Half of
the galaxies with embedded structures are found in this area but some of
them could have a distance to the cluster centre larger than their
projected distance. In the bottom panel of Fig.~\ref{figdproj}, the ratio
of $\alpha$-elements over iron in the transition area is not as large as
for the innermost part of the cluster and spreads over the same range as
for galaxies at larger projected distances.

In their study of Virgo early-type dwarf galaxies, Lisker et al. (2007
and references therein) do not give the stellar population properties
of the three subclasses, which are the result of infalling galaxy
transformation. If we put together the eight dwarfs corresponding to
their dE(di) subclass and the five other dwarfs, which exhibit very
young central stellar populations, this subsample presents the same
spatial and velocity distributions as their least unrelaxed dwarf
galaxy population. The velocity histogram is more spread for the 13
selected dwarfs compared to that of the 24 remaining early-type
dwarfs, including the two galaxies for which we did not succeed
to fit the spectra due to low signal-to-noise ratios.
If we also take into account their corresponding spatial
distributions, we can conclude that the subclass is probably not fully
relaxed.  A clear difference exists in the stellar population ages,
since among the 22 dwarfs with no embedded structures,
only six galaxies have ages between 3 and
9~Gyr.

The brighter part of our sample can be compared to existing studies of
stellar populations of early-type galaxies. Poggianti et al. (2001)
presented the luminosity-weighted ages and metallicities for several
dozens of Coma cluster early-type galaxies. From their Fig.~2, it is
clear that there are two populations of galaxies, old and young. The
age of the old ones for $M_B > -19.0$ anticorrelates with the
luminosity. In our sample, however, we do not have any bright galaxy
with a young stellar population. This is probably due to the selection
of objects in our sample: (1) we tried to avoid galaxies with redshifts
already in the literature; (2) young galaxies have higher surface
brightnesses compared to old ones for a given size, therefore, they have
a better chance to have already been observed in redshift surveys with
limited aperture magnitudes, and thus to be excluded from our sample.

S\'anchez et al. (2007) have obtained internal kinematics and stellar
population parameters for galaxies in the core of the Abell~2218
cluster using 3D-spectroscopy of the cluster centre. The behavior of
age estimates is similar to that of the \object{Abell~496} galaxies: for
low-mass objects, ages tend to be younger and more spread out than for
large galaxies.

\section{Discussion}

\subsection{On the origin of dE/dS0 galaxies}

We discuss here what we can learn from the stellar population
properties of dE/dS0s in \object{Abell~496} regarding their origin and
evolution.  First, what is the explanation for the spread in the
stellar ages and metallicities of our dwarf galaxy sample? Is this
spread related to some differences in the time needed for galaxies to
be accreted into the cluster and to reach the cluster core? Low-mass
galaxies are more sensitive than massive galaxies to both internal and
external processes affecting their gas contents and star formation
rates. Therefore, the ``in situ'' formation scenario, which is able to
reproduce the various timescales for star formation, could work as
well as an accretion scenario. Which scenario can explain that star
formation occurs in the nuclear regions for a number of dwarfs? Star
formation is driven in particular by the gas content, and if the gas is
rapidly removed from its host galaxy, star formation can be
stopped in a very short time. On the other hand, an infall of gas in
the galaxy centre will induce star formation in the nuclear
region. The scenario should also reproduce the observed [Mg/Fe] ratio
and its relation with the velocity dispersion.

Three possible scenarii for the gas removal usually considered for dE
galaxies are: (1) supernova-driven winds at the early stages of galaxy
evolution (Dekel \& Silk 1986); (2)~ram pressure stripping by the
intra-cluster medium (e.g., Marcolini et al. 2003); and (3)~tidal
harassment due to distant and repeated encounters with other cluster
members (Moore et al.  1998). The first one is often referred to as
``internal'' and latter two as ``external'' agents of dE galaxy
formation and evolution.

The idea of supernova-driven winds is based on the assumption that the
gravitational field of dwarf galaxies is not sufficiently strong to keep the
interstellar medium from being swept out by SN~II explosions during the
first intense star formation episode.

Several different models for galactic winds exist (see De Rijcke
et al.  2005 for a detailed review). Simple models discussed in Yoshii
\& Arimoto (1987) lead to abrupt gas loss and interruption of the star
formation episode after a short time ($10^7$ years). Consequently,
later explosions of SNIa on a timescale of gigayears will not
contribute to the iron enrichment of the stellar population 
(Matteucci 1994). Thus, a short star formation episode will lead to an
overabundance of $\alpha$-elements over iron ([$\alpha$/Fe] $>$
0). This phenomenon is observed in globular clusters and, usually, in
giant early-type galaxies (Kuntschner et al. 2006, Sil'chenko
2006). If gas is removed from dE galaxies by supernova-driven winds,
we would expect to see [Mg/Fe]$>$0~dex, and it should anticorrelate
with the dynamical mass of galaxies (or $\sigma$). This
does not agree with what we see in our data.

Another study by Chiosi \& Carraro (2002), contrary to Yoshii \& Arimoto
(1987), predicts very long and oscillating star formation histories in
dwarf galaxies: supernova explosions disperse the gas, stopping star
formation, but later the gas cools down, falls back in and another
star-formation episode begins. In this case, the [Mg/Fe] ratios decrease
to $\sim$0~dex, the correlation between the dynamical mass and
metallicity can be explained by the lower efficiency of star formation
in galaxies of lower masses due to more efficient supernova
feedback. However, under this scenario, one would not observe any old
dwarf ellipticals because younger stars, formed in the secondary
episodes of star formation, would dominate the light even though their
masses might be low compared to the masses of the first generation of
stars. Therefore, one still needs a mechanism to sweep out the remaining
gas from the low-mass galaxies earlier or later during their lifetime.

In the case of \object{Abell~496}, [Mg/Fe]$\approx$0~dex for nearly all low-mass objects
($\sigma_0<$60 km/s), meaning that the star formation epoch durations were
at least 1$-$2 Gyr, the minimal required time to complete the iron
enrichment (Matteucci 1994). Hence, we cannot consider the scenario of gas
removal by supernova-driven winds as the only explanation for the observed
properties of dE galaxies, although it allows us to reproduce the
observed mass-metallicity correlation.

Ram-pressure stripping of late-type dwarf galaxies (dIrr) or dwarf
spirals appears to be an acceptable way to remove the gas from dE
progenitors.  If we assume that late-type galaxies have formed outside
the central region of the cluster, and later fell onto it, sufficient
time is left for iron enrichment, since the typical infall time is a few
Gyr. If ram-pressure stripping plays the leading role in gas removal,
one would expect a large spread of luminosity-weighted ages for low-mass
objects that can be completely stripped during their first passage
through the cluster centre.  This stripping can occur at any moment
during the galaxy lifetime. Due to our small sample we cannot give a
decisive answer to whether the spread of age estimations in
Fig.~\ref{figtZsig} is the result of ram-pressure stripping of late-type
progenitors, or due to the low quality of measurements (including
the age-metallicity degeneracy effects).

\begin{figure}
\includegraphics[width=8cm]{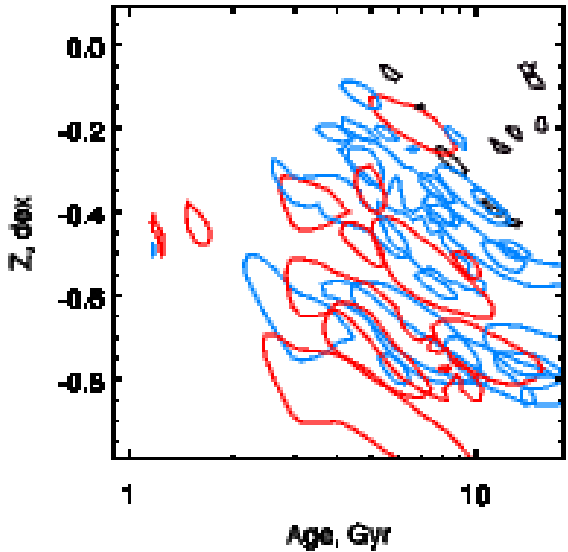}
\caption{1-$\sigma$ confidence levels of age and metallicity determinations
for the \object{Abell~496} galaxies. Objects in the three velocity dispersion bins
are plotted using different colours: red for $\sigma < 30$~km~s$^{-1}$, 
blue for $30 < \sigma < 60$~km~s$^{-1}$, and black for 
$\sigma > 60$~km~s$^{-1}$.\label{figtzell}}
\end{figure}

Smith et al. (2007) pointed out that there is a conspiracy in the sense
that the distribution of galaxies in the $Z - \log t$ plane at fixed
velocity dispersion is aligned with the age-metallicity degeneracy.  In
Fig.\ref{figtzell}, we plotted the relationship between age and
metallicity for different intervals of sigma with the Smith et
al. sample of faint red galaxies in three clusters located inside the
Shapley Supercluster. We choose two intervals in the same range and
separately present the galaxies where sigma is lower than 30~km~s$^{-1}$
(not found in the Smith et al. survey). For the two intervals we had in
common with the Smith et al.  survey, we confirm the existence of the
age-metallicity anticorrelation for fixed velocity dispersion. The
spread in age is not due to the age-metallicity degeneracy since the
measurement errors account for less than one fourth of the dispersion
when $\sigma >$ 30~km~s$^{-1}$. For lower velocity dispersions the
spread in metallicity is clearly not due to degeneracy and the SSP model
cannot be applied if star formation occurs at low rate, but during a
significant period of time. Note the peculiar location of four galaxies
with high sigma values, high metallicities, and old ages.

The existence of cores of various colours in a number of dE/dS0 galaxies
of our sample are related to the young nuclei in dE galaxies, which were
recently discovered in three dwarfs in the Virgo cluster (Chilingarian
et al. 2007b) and in a low-luminosity S0 in a group (Chilingarian et
al. 2008b).  They proposed that ram-pressure stripping is
a way to expel the ISM from the outer parts of a low-mass disc galaxy
(dE progenitor), and, at the same time, to compress the gas and to
induce star formation in the dense nuclear region.  Depending on the
time and duration of this phenomenon, the colour of the nuclear region
can be blue (when the star formation event is still going on or has
finished recently) or red (if the star formation period ended long ago,
so only the metallicity excess can be easily detected). Although the
quantitative modelling of this phenomenon is quite complex and requires
extensive numerical simulations, qualitatively, we consider the
presence of cores of various colours in our sample of dE galaxies as an
argument for the ram-pressure stripping scenario.

The presence of faint embedded discs in some galaxies is another strong
argument for an evolutionary connection between early- and late- type dwarf
galaxies. This result is in agreement with N-body modelling of morphological
evolution of late-type galaxies in clusters (Mastropietro et al. 2005),
suggesting that discs will not be completely destroyed. This is supported by
the fact that embedded structures are revealed only in brighter dE galaxies,
which are more resistant to tidal harassment.

From the correlations among early-type galaxies in the SDSS Clemens et
al.  (2006) found that the metallicity grows monotonically with $\sigma$
and interpreted this correlation as due to the stellar mass build-up
being regulated by the halo mass. We confirm this trend, which was also
noticed by other studies, but find that the environment plays an
important role for suppressing star formation in dwarf ellipticals and
causing morphological transformation as revealed in nearby clusters
(e.g., Michielsen et al., 2008).

Our main conclusion is that dE galaxies have late-type progenitors
and have formed in the peripheral parts of the cluster, experiencing
tidal interactions with the cluster potential and other cluster
galaxies, as well as ram-pressure stripping while crossing the cluster
centre.

\subsection{Galaxies in a dense environment}

There are five galaxies belonging to \object{Abell~496} located above the bulk
of objects in the Faber-Jackson relation (Fig.~\ref{figfjr}).  Four of them
(ACO496J043333.17-131712.6 (G-17); ACO496J043341.69-131551.8 (G-24);
ACO496J043337.35-131520.2 (G-20); ACO496J043338.22-131500.7 (G-21)) are located
within 100~kpc of the cD in projected distance, while the fifth one
(ACO496J043401.57-131359.7 (G-40)) is located at 240~kpc. All five objects
have a number of common properties: (1)~old stellar populations
($>$12~Gyr); (2)~very high Mg abundances (Mg$b > 4.2$~\AA) and [Mg/Fe]
abundance ratios (0.18 $<$ [Mg/Fe] $<$ 0.45~dex); (3)~high surface
brightnesses and structural properties in the continuity of those of
normal ellipticals and bulges of spirals (Fig.~\ref{figmuav}). These objects are
all ellipticals and show high surface brightness red cores in their
colour maps and no embedded structures.  Four of them are notably
above the Mg$b-\sigma$ correlation from Smith et al.  (2006), while
the fifth one (ACO496J043401.57-131359.7 (G-40)) falls exactly on the
correlation.  The latter object is the one located 240~kpc from the cD
and is the only elongated one; its properties are therefore not as
extreme as those of the four other galaxies.

As evidenced in Fig.~\ref{figdproj}, we note the complete avoidance of
the inner region of the cluster by young galaxies.  That \object{Abell~496} is a
dynamically relaxed cluster explains this effect: all the galaxies we
observe now in the central region of the cluster have been captured a
long time ago by the cluster potential.

We translate the [Mg/Fe] ratio into duration of the starburst using
prescriptions from Thomas et al. (2005). According to their formula~4,
[Mg/Fe]=0.45~dex corresponds to 30~Myr and [Mg/Fe]=0.3~dex to
250~Myr. The four objects described above have high Mg abundances for
their velocity dispersions (which are already too high for their
luminosities), but normal iron abundances and average
metallicities. This suggests that when Mg was produced these galaxies
were more massive than now, and environmental effects have occurred
before the production of iron, i.e. during the first 1~Gyr of their
evolution.  In view of their proximity to the cD, such properties can be
explained by tidal stripping during their infall onto the cluster
centre.  The most extreme cases are compact ellipticals such as
ACO496J043337.35-131520.2 (G-20) that we discovered in this cluster
(Chilingarian et al.  2007c), which may have lost up to 90\% of its
stellar mass. The progenitors of the three other galaxies have undergone
stellar mass loss as well, although not as strong. If we combine all the
information derived from the kinematics and stellar population analyses,
we estimate the progenitor luminosities to range between
$M_B\approx-19$~mag for cE (G-20) and $M_B\approx-21$~mag for
ACO496J043341.69-131551.8 (G-24).  The similar process applied to
fainter early-type galaxies (nucleated dE's or dS0's) may lead to the
formation of ultra-compact dwarfs (Bekki et al. 2003) or UCD/cE
transitional objects, as the one found by Chilingarian \& Mamon (2008).
Another unusual galaxy in the cluster centre (54~kpc from the cD in
projection) is ACO496J043332.07-131518.1 (G-16): it has an extreme value
of [Mg/Fe] (+0.28~dex) for its velocity dispersion ($79\pm
1$~km~s$^{-1}$) and luminosity ($M_B=-17.91$~mag). Taking into account
its old population (14~Gyr), it probably has an origin similar to
that of the four galaxies discussed above.

Objects close to the cluster centre are, therefore, strongly affected by
their high-density environment in the central part of the cluster. We
can make a rough estimate for the lower limit of the total dynamical
mass in the centre of the cluster by taking the radial velocity of
ACO496J043333.53-131852.6 (G-18) and assuming it is on a circular orbit
observed perpendicular to the line of sight with a radius of
$\approx$130~kpc. This will result in a value of $M \approx \Delta v^2
r_{proj} / 2G = 4 \cdot 10^{13} M_{\odot}$ that is consistent with the
total mass estimate from the analysis of X-ray data, which gives $\sim3
\cdot 10^{13} M_{\odot}$ within 150~kpc (T. F. Lagana et al.  in
preparation).

\section{Conclusions}

We have conducted an analysis of spectral and photometric data for a
unique sample of dwarf early-type galaxies extending toward small
objects having low velocity dispersions, down to $\sigma \sim
20$~km~s~$^{-1}$.

Although dwarf elliptical and lenticular galaxies with shallow brightness
profiles are structurally different from the ``classical'' elliptical
galaxies and bulges of spirals forming the Kormendy relation, their stellar 
population properties seem to follow and extend known correlations of
absorption line strengths, metallicity, and age versus $\sigma$.

Low-mass galaxies ($\sigma <70$~km~s~$^{-1}$) show solar [Mg/Fe]
element abundance ratios, arguing for long durations of the star
formation episodes ($>\sim$1.5~Gyr)

Based on the properties of the observed galaxies, we conclude that there
must be an evolutionary connection between late- and early-type dwarf
galaxies; external agents (ram-pressure stripping, harassment, etc.) must
play a key role in the morphological transformation.

Evolution of even more massive galaxies residing in the central part of the
cluster is ruled by environmental effects: tidal stripping of stellar
discs by the cluster potential is one of the probable scenarii of formation
and evolution of relatively compact and dense elliptical galaxies, observed
only in the vicinity of the cD.

\begin{acknowledgements}
We acknowledge support for IC's PhD thesis in 2005--2006 by the INTAS foundation
(project 04-83-3618). We are very grateful to the organizing committee
of the conference ``Mapping the Galaxy and Nearby Galaxies'', who
provided IC with the financial support essential to attend the meeting
and present preliminary results of this study. Additional support is
given by the RFBR-Flaanders project 05-02-19805 on studies of dwarf
elliptical galaxies (co-PI: O.  Sil'chenko). Special thanks to the
VO-Paris project (M.-L.  Dubernet, P. Le Sidaner) for funding several
short missions of IC to Paris during 2005--2006. We are grateful to
the staff of the Terapix data centre at IAP, France, for their
efficiency and competence in reducing our Megacam imaging data and to
C.~Marchais and N.~Bavouzet for their help during the early stages of
the spectroscopic analysis. This research has made use of (1) the
NASA/IPAC Extragalactic Database (NED) which is operated by the Jet
Propulsion Laboratory, California Institute of Technology, under
contract with the National Aeronautics and Space Administration; (2)
Aladin developed by the Centre de Donn\'ees astronomiques de
Strasbourg; (3) SAOImage DS9, developed by Smithsonian Astrophysical
Observatory; (4) XMM Newton Science archive, operated by the European
Space Agency. Special thanks to: A.~Graham for critical reading of the
manuscript and fruitful discussions, M.~Koleva for useful suggestions
regarding the spectral fitting technique; P.~S\'anchez-Bl\'azquez for
providing stellar population parameters for early-type galaxies in a
computer-readable format; E.~Slezak for making available to us his
wavelet analysis software.  We thank our anonymous referee for
constructive comments.
\end{acknowledgements}

\appendix

\section{Atlas of 46 members of \object{Abell~496}.}

\begin{table*}
\caption{$B$ band absolute magnitudes, morphological classification, 
detection of embedded structures, 
and total magnitudes in 4 CFHT bands for
48 \object{Abell~496} galaxies, including 2 objects marked A1 and A2 for which no
spectral fit was possible. 
\label{tabmorph}}
\begin{tabular}{llcllcccc}
\hline
\hline
N & IAU Name & $M_B$ & type & emb.str. & $u^*$ & $g'$ & $r'$ & $i'$ \\
  &          & mag &   &     & mag & mag & mag & mag \\
\hline
01 & ACO496J043306.97-131238.8 & -17.72 & SB0a/dS0 & (s):B+S & 19.09$\pm$0.03 & 17.67$\pm$0.04 & 17.08$\pm$0.04 & 16.68$\pm$0.04 \\
02 & ACO496J043308.85-130235.6 & -18.54 & Sbc & (s):D & 18.23$\pm$0.02 & 16.85$\pm$0.03 & 16.23$\pm$0.03 & 15.85$\pm$0.03 \\
03 & ACO496J043312.08-130449.3 & -16.55 & dE & - & 20.12$\pm$0.12 & 18.84$\pm$0.14 & 18.35$\pm$0.06 & 17.96$\pm$0.07 \\
04 & ACO496J043317.75-131536.6 & -16.58 & dS0/dE & (w):S & 19.93$\pm$0.17 & 18.81$\pm$0.06 & 18.28$\pm$0.06 & 17.96$\pm$0.04 \\
05 & ACO496J043318.95-131726.9 & -16.99 & dE & - & 19.81$\pm$0.05 & 18.40$\pm$0.03 & 17.96$\pm$0.06 & 17.66$\pm$0.06 \\
06 & ACO496J043320.35-130314.9 & -18.78 & SB0 & (s):B/S/R? & 18.01$\pm$0.02 & 16.61$\pm$0.03 & 16.00$\pm$0.03 & 15.62$\pm$0.03 \\
07 & ACO496J043321.37-130416.6 & -17.01 & dSB0a & (s):B/S/R? & 19.38$\pm$0.07 & 18.38$\pm$0.03 & 17.97$\pm$0.05 & 17.71$\pm$0.04 \\
08 & ACO496J043324.61-131111.9 & -15.69 & dE & - & 21.08$\pm$0.08 & 19.70$\pm$0.09 & 19.13$\pm$0.06 & 18.78$\pm$0.08 \\
09 & ACO496J043324.91-131342.6 & -17.20 & dE & - & 19.57$\pm$0.05 & 18.19$\pm$0.04 & 17.60$\pm$0.04 & 17.25$\pm$0.05 \\
10 & ACO496J043325.10-130906.6 & -15.93 & dE & - & 20.66$\pm$0.09 & 19.46$\pm$0.04 & 18.87$\pm$0.15 & 18.63$\pm$0.07 \\
11 & ACO496J043325.40-131414.6 & -16.66 & dE & - & 20.06$\pm$0.04 & 18.73$\pm$0.07 & 18.17$\pm$0.04 & 17.82$\pm$0.04 \\
12 & ACO496J043325.54-130408.0 & -16.07 & dE & - & 20.60$\pm$0.07 & 19.32$\pm$0.04 & 18.90$\pm$0.07 & 18.62$\pm$0.07 \\
13 & ACO496J043326.49-131717.8 & -17.13 & dS0  & - & 19.63$\pm$0.03 & 18.26$\pm$0.05 & 17.72$\pm$0.05 & 17.35$\pm$0.06 \\
14 & ACO496J043329.79-130851.7 & -16.49 & dE & - & 20.03$\pm$0.05 & 18.90$\pm$0.07 & 18.39$\pm$0.06 & 18.05$\pm$0.04 \\
15 & ACO496J043331.48-131654.6 & -18.13 & S0a & (s):S/D & 18.67$\pm$0.05 & 17.26$\pm$0.04 & 16.66$\pm$0.03 & 16.32$\pm$0.04 \\
16 & ACO496J043332.07-131518.1 & -17.76 & dE/E/S0 & (w):R & 19.11$\pm$0.03 & 17.63$\pm$0.05 & 16.96$\pm$0.04 & 16.53$\pm$0.03 \\
17 & ACO496J043333.17-131712.6 & -18.27 & E & - & 18.68$\pm$0.04 & 17.12$\pm$0.03 & 16.46$\pm$0.04 & 16.05$\pm$0.04 \\
18 & ACO496J043333.53-131852.6 & -18.17 & SB0a & (s):B & 18.62$\pm$0.03 & 17.22$\pm$0.03 & 16.60$\pm$0.03 & 16.22$\pm$0.03 \\
19 & ACO496J043334.54-131137.1 & -16.04 & dE & - & 20.65$\pm$0.15 & 19.35$\pm$0.13 & 18.85$\pm$0.04 & 18.50$\pm$0.07 \\
20 & ACO496J043337.35-131520.2 & -16.99 & cE & - & 19.93$\pm$0.03 & 18.40$\pm$0.06 & 17.60$\pm$0.06 & 17.20$\pm$0.08 \\
21 & ACO496J043338.22-131500.7 & -17.44 & dE & - & 19.67$\pm$0.02 & 17.95$\pm$0.04 & 17.26$\pm$0.05 & 16.81$\pm$0.04 \\
22 & ACO496J043339.07-131319.7 & -15.63 & dE & - & 21.22$\pm$0.20 & 19.76$\pm$0.10 & 19.06$\pm$0.10 & 18.70$\pm$0.11 \\
23 & ACO496J043339.72-131424.6 & -16.17 & dE & - & 20.59$\pm$0.04 & 19.22$\pm$0.03 & 18.63$\pm$0.05 & 18.24$\pm$0.10 \\
24 & ACO496J043341.69-131551.8 & -18.49 & E & - & 18.54$\pm$0.04 & 16.90$\pm$0.04 & 16.17$\pm$0.04 & 15.71$\pm$0.03 \\
25 & ACO496J043342.10-131653.7 & -17.65 & dS0/S0 & (s):D/R & 19.28$\pm$0.05 & 17.74$\pm$0.04 & 17.07$\pm$0.04 & 16.62$\pm$0.04 \\
26 & ACO496J043342.13-131258.8 & -16.92 & dE & - & 19.97$\pm$0.06 & 18.47$\pm$0.04 & 17.84$\pm$0.02 & 17.53$\pm$0.02 \\
27 & ACO496J043342.83-130846.8 & -17.92 & S0/E & - & 18.81$\pm$0.03 & 17.47$\pm$0.04 & 16.84$\pm$0.04 & 16.42$\pm$0.04 \\
28 & ACO496J043343.04-130514.1 & -17.41 & dS0 & - & 19.33$\pm$0.03 & 17.98$\pm$0.03 & 17.36$\pm$0.03 & 16.95$\pm$0.04 \\
29 & ACO496J043343.04-125924.4 & -16.34 & dS0/dE & - & 20.35$\pm$0.14 & 19.05$\pm$0.07 & 18.55$\pm$0.04 & 18.21$\pm$0.27 \\
30 & ACO496J043345.67-130542.2 & -17.32 & dS0 & - & 19.39$\pm$0.06 & 18.07$\pm$0.06 & 17.53$\pm$0.03 & 17.14$\pm$0.06 \\
31 & ACO496J043346.71-131756.2 & -18.44 & SB0 & (s):S+B & 18.42$\pm$0.04 & 16.95$\pm$0.04 & 16.34$\pm$0.03 & 15.96$\pm$0.03 \\
32 & ACO496J043348.59-130558.3 & -17.09 & dE & - & 19.71$\pm$0.06 & 18.30$\pm$0.04 & 17.69$\pm$0.03 & 17.35$\pm$0.07 \\
33 & ACO496J043349.08-130520.5 & -17.96 & S0a/dS0 & (w):S & 18.83$\pm$0.02 & 17.43$\pm$0.03 & 16.85$\pm$0.04 & 16.44$\pm$0.04 \\
34 & ACO496J043350.17-125945.4 & -16.08 & dS0 & - & 20.56$\pm$0.11 & 19.31$\pm$0.09 & 18.79$\pm$0.10 & 18.44$\pm$0.09 \\
35 & ACO496J043351.54-131135.5 & -16.47 & dS0/dE & - & 20.17$\pm$0.05 & 18.92$\pm$0.10 & 18.43$\pm$0.09 & 18.23$\pm$0.08 \\
36 & ACO496J043352.77-131523.8 & -17.73 & S0/dS0 & (w):D? & 18.94$\pm$0.06 & 17.66$\pm$0.04 & 17.08$\pm$0.05 & 16.69$\pm$0.04 \\
37 & ACO496J043355.55-131024.9 & -16.12 & dS0/dE & - & 20.50$\pm$0.13 & 19.27$\pm$0.07 & 18.68$\pm$0.07 & 18.40$\pm$0.06 \\
38 & ACO496J043356.18-125913.1 & -17.96 & E/dE/dS0 & - & 18.52$\pm$0.02 & 17.43$\pm$0.04 & 16.95$\pm$0.03 & 16.60$\pm$0.03 \\
39 & ACO496J043359.03-130626.7 & -17.50 & dS0 & - & 19.17$\pm$0.03 & 17.89$\pm$0.04 & 17.34$\pm$0.04 & 16.98$\pm$0.04 \\
40 & ACO496J043401.57-131359.7 & -18.34 & E/S0 & - & 18.44$\pm$0.03 & 17.05$\pm$0.03 & 16.41$\pm$0.03 & 16.04$\pm$0.03 \\
41 & ACO496J043403.19-131310.6 & -18.75 & SB0/SBa & (s):S+B & 18.14$\pm$0.02 & 16.64$\pm$0.03 & 16.00$\pm$0.04 & 15.62$\pm$0.04 \\
42 & ACO496J043408.50-131152.7 & -16.67 & dE/dS0 & (w):B?/D & 19.97$\pm$0.05 & 18.72$\pm$0.12 & 18.13$\pm$0.10 & 17.80$\pm$0.14 \\
43 & ACO496J043410.60-130756.7 & -16.79 & dS0 & - & 19.89$\pm$0.05 & 18.60$\pm$0.07 & 18.07$\pm$0.03 & 17.74$\pm$0.06 \\
44 & ACO496J043413.00-131003.5 & -18.25 & E/S0 & - & 18.61$\pm$0.02 & 17.14$\pm$0.03 & 16.54$\pm$0.03 & 16.12$\pm$0.03 \\
45 & ACO496J043413.08-131231.6 & -16.60 & dE & - & 20.14$\pm$0.09 & 18.79$\pm$0.03 & 18.24$\pm$0.09 & 17.91$\pm$0.10 \\
46 & ACO496J043415.37-130823.5 & -17.46 & dIm & - & 18.49$\pm$0.03 & 17.93$\pm$0.05 & 17.81$\pm$0.06 & 17.67$\pm$0.06 \\
\hline
A1 & ACO496J043411.72-131130.2 & -15.07 & dE & - & 21.53$\pm$0.09 & 20.32$\pm$0.09 & 19.78$\pm$0.08 & 19.47$\pm$0.08 \\
A2 & ACO496J043414.54-131303.0 & -15.50 & dE & - & 20.87$\pm$0.19 & 19.89$\pm$0.11 & 19.32$\pm$0.17 & 19.00$\pm$0.17 \\
\hline
\hline
\end{tabular}
\end{table*}

\begin{table*}
\caption{Projected distances, relative radial velocities with respect to the 
cluster centre, and structural parameters: effective radii and mean surface 
brightnesses in $B$ within 1~$r_{eff}$, radial velocities, velocity
dispersions, SSP-equivalent ages and metallicities for 46 \object{Abell~496} galaxies.
Columns 7 -- 9 contain best-fitting parameters, while 10 -- 12 ones, 
obtained with the scan of the $\chi^2$ values in the age-metallicity space.
\label{tabfit}}
\begin{tabular}{lcccccccc}
\hline
\hline
N & $d_{proj}$ & $v_r - v_{A496}$ & $r_{eff}$ & $\langle\mu_{eff}\rangle$ & $v$ & $\sigma_{fit}$ & $t_{fit}$ & $Z_{fit}$\\ 
  & arcsec     & km~s$^{-1}$  & arcsec & mag~arcsec$^{-2}$ & km~s$^{-1}$ & km~s$^{-1}$ & Gyr & dex \\ 
\hline
01 & 486 & -970 & 2.94$\pm$0.12 & 22.31$\pm$0.04 & 8915$\pm$1 & 47$\pm$1 & 4.9$\pm$0.9 & -0.13$\pm$0.03 \\
02 & 892 & 977 & 6.20$\pm$0.14 & 23.11$\pm$0.05 & 10862$\pm$1 & 45$\pm$1 & 4.0$\pm$0.6 & -0.21$\pm$0.04 \\
03 & 752 & -691 & 2.52$\pm$0.34 & 23.16$\pm$0.12 & 9194$\pm$3 & 33$\pm$3 & 14.2$\pm$6.7 & -0.73$\pm$0.14 \\
04 & 293 & 1179 & 3.07$\pm$0.16 & 23.55$\pm$0.09 & 11064$\pm$2 & 18$\pm$3 & 4.9$\pm$3.1 & -0.72$\pm$0.09 \\
05 & 295 & -1260 & 2.18$\pm$0.10 & 22.39$\pm$0.06 & 8625$\pm$2 & 49$\pm$2 & 14.0$\pm$4.1 & -0.54$\pm$0.10 \\
06 & 788 & -1120 & 3.38$\pm$0.12 & 21.56$\pm$0.05 & 8765$\pm$1 & 72$\pm$1 & 6.8$\pm$0.4 & -0.14$\pm$0.01 \\
07 & 725 & -2149 & 1.94$\pm$0.06 & 22.13$\pm$0.06 & 7736$\pm$1 & 27$\pm$1 & 1.2$\pm$0.2 & -0.46$\pm$0.03 \\
08 & 331 & -292 & 1.41$\pm$0.10 & 22.75$\pm$0.07 & 9593$\pm$2 & 21$\pm$2 & 9.9$\pm$4.9 & -0.54$\pm$0.07 \\
09 & 222 & -686 & 2.13$\pm$0.08 & 22.13$\pm$0.05 & 9199$\pm$1 & 41$\pm$1 & 12.0$\pm$3.0 & -0.51$\pm$0.05 \\
10 & 436 & 528 & 2.03$\pm$0.06 & 23.31$\pm$0.08 & 10413$\pm$3 & 26$\pm$3 & 6.2$\pm$4.6 & -0.72$\pm$0.12 \\
11 & 201 & -1510 & 1.75$\pm$0.14 & 22.26$\pm$0.08 & 8375$\pm$1 & 39$\pm$1 & 9.5$\pm$2.6 & -0.35$\pm$0.03 \\
12 & 716 & 1208 & 2.30$\pm$0.13 & 23.43$\pm$0.07 & 11093$\pm$3 & 26$\pm$3 & 5.5$\pm$2.5 & -0.89$\pm$0.07 \\
13 & 192 & -944 & 4.15$\pm$0.28 & 23.65$\pm$0.06 & 8941$\pm$2 & 40$\pm$2 & 6.6$\pm$2.3 & -0.37$\pm$0.06 \\
14 & 426 & -1244 & 1.68$\pm$0.11 & 22.33$\pm$0.11 & 8641$\pm$2 & 31$\pm$2 & 6.6$\pm$2.4 & -0.72$\pm$0.06 \\
15 & 118 & -298 & 2.85$\pm$0.10 & 21.84$\pm$0.05 & 9587$\pm$1 & 51$\pm$1 & 5.5$\pm$0.5 & -0.21$\pm$0.01 \\
16 & 87 & 69 & 1.69$\pm$0.09 & 21.07$\pm$0.10 & 9954$\pm$1 & 79$\pm$1 & 12.9$\pm$1.4 & -0.43$\pm$0.03 \\
17 & 114 & -13 & 1.51$\pm$0.05 & 20.34$\pm$0.06 & 9872$\pm$1 & 124$\pm$2 & 14.0$\pm$1.2 & -0.06$\pm$0.02 \\
18 & 202 & 1820 & 2.34$\pm$0.11 & 21.38$\pm$0.06 & 11705$\pm$1 & 83$\pm$1 & 5.7$\pm$0.5 & -0.07$\pm$0.01 \\
19 & 248 & -1441 & 2.25$\pm$0.35 & 23.42$\pm$0.18 & 8444$\pm$2 & 19$\pm$3 & 6.1$\pm$4.3 & -0.50$\pm$0.11 \\
20 & 22 & -130 & 0.75$\pm$0.02 & 20.07$\pm$0.07 & 9755$\pm$1 & 103$\pm$1 & 15.7$\pm$1.2 & -0.06$\pm$0.02 \\
21 & 41 & 407 & 1.22$\pm$0.05 & 20.69$\pm$0.05 & 10292$\pm$2 & 147$\pm$2 & 15.6$\pm$1.9 & -0.18$\pm$0.03 \\
22 & 142 & 929 & 1.51$\pm$0.13 & 22.95$\pm$0.09 & 10814$\pm$1 & 26$\pm$2 & 6.9$\pm$1.9 & -0.20$\pm$0.06 \\
23 & 81 & 256 & 1.74$\pm$0.04 & 22.72$\pm$0.07 & 10141$\pm$2 & 37$\pm$3 & 15.9$\pm$6.5 & -0.82$\pm$0.14 \\
24 & 58 & -114 & 1.37$\pm$0.06 & 19.87$\pm$0.07 & 9771$\pm$1 & 176$\pm$1 & 14.8$\pm$0.9 & -0.09$\pm$0.01 \\
25 & 96 & -369 & 2.51$\pm$0.11 & 22.04$\pm$0.06 & 9516$\pm$1 & 77$\pm$1 & 8.5$\pm$1.2 & -0.28$\pm$0.02 \\
26 & 174 & 463 & 4.44$\pm$0.18 & 24.01$\pm$0.08 & 10348$\pm$3 & 36$\pm$3 & 6.0$\pm$3.5 & -0.41$\pm$0.10 \\
27 & 420 & 649 & 2.84$\pm$0.12 & 22.04$\pm$0.04 & 10534$\pm$1 & 53$\pm$1 & 7.7$\pm$1.1 & -0.58$\pm$0.02 \\
28 & 631 & -203 & 2.64$\pm$0.11 & 22.40$\pm$0.06 & 9682$\pm$1 & 56$\pm$1 & 12.9$\pm$2.3 & -0.49$\pm$0.04 \\
29 & 979 & 338 & 3.02$\pm$0.19 & 23.76$\pm$0.09 & 10223$\pm$2 & 21$\pm$2 & 11.2$\pm$5.4 & -0.75$\pm$0.11 \\
30 & 610 & -196 & 2.39$\pm$0.22 & 22.27$\pm$0.06 & 9689$\pm$1 & 46$\pm$1 & 4.7$\pm$0.6 & -0.21$\pm$0.03 \\
31 & 188 & -1519 & 2.84$\pm$0.10 & 21.53$\pm$0.05 & 8366$\pm$1 & 73$\pm$1 & 11.6$\pm$1.2 & -0.40$\pm$0.02 \\
32 & 603 & -108 & 2.03$\pm$0.12 & 22.15$\pm$0.08 & 9777$\pm$1 & 46$\pm$1 & 10.9$\pm$1.6 & -0.38$\pm$0.02 \\
33 & 642 & -115 & 3.66$\pm$0.18 & 22.55$\pm$0.05 & 9770$\pm$1 & 55$\pm$1 & 7.9$\pm$1.4 & -0.39$\pm$0.02 \\
34 & 972 & 492 & 2.48$\pm$0.18 & 23.59$\pm$0.11 & 10377$\pm$3 & 39$\pm$4 & 3.0$\pm$1.7 & -0.67$\pm$0.21 \\
35 & 317 & -426 & 1.71$\pm$0.21 & 22.38$\pm$0.13 & 9459$\pm$1 & 21$\pm$2 & 1.5$\pm$0.4 & -0.43$\pm$0.03 \\
36 & 219 & -959 & 2.52$\pm$0.12 & 21.97$\pm$0.06 & 8926$\pm$1 & 52$\pm$1 & 7.8$\pm$1.3 & -0.34$\pm$0.02 \\
37 & 409 & -1459 & 2.25$\pm$0.13 & 23.34$\pm$0.09 & 8426$\pm$2 & 30$\pm$2 & 5.4$\pm$2.1 & -0.29$\pm$0.05 \\
38 & 1024 & 1318 & 1.71$\pm$0.09 & 20.90$\pm$0.04 & 11203$\pm$1 & 43$\pm$1 & 1.2$\pm$0.1 & -0.47$\pm$0.02 \\
39 & 635 & 667 & 4.46$\pm$0.21 & 23.45$\pm$0.06 & 10552$\pm$1 & 27$\pm$2 & 4.9$\pm$1.4 & -0.48$\pm$0.04 \\
40 & 361 & 398 & 1.34$\pm$0.07 & 20.01$\pm$0.05 & 10283$\pm$1 & 151$\pm$1 & 12.9$\pm$0.9 & -0.21$\pm$0.01 \\
41 & 400 & -933 & 3.41$\pm$0.11 & 21.61$\pm$0.05 & 8952$\pm$1 & 87$\pm$1 & 11.9$\pm$1.1 & -0.25$\pm$0.02 \\
42 & 503 & -123 & 2.83$\pm$0.30 & 23.29$\pm$0.18 & 9762$\pm$3 & 31$\pm$4 & 6.7$\pm$5.0 & -0.70$\pm$0.11 \\
43 & 667 & -1510 & 2.90$\pm$0.24 & 23.22$\pm$0.06 & 8375$\pm$2 & 38$\pm$2 & 4.0$\pm$1.6 & -0.36$\pm$0.09 \\
44 & 615 & 950 & 3.27$\pm$0.11 & 22.02$\pm$0.04 & 10835$\pm$1 & 50$\pm$1 & 8.9$\pm$1.1 & -0.23$\pm$0.02 \\
45 & 549 & 314 & 2.02$\pm$0.06 & 22.62$\pm$0.06 & 10199$\pm$2 & 43$\pm$3 & 14.1$\pm$5.4 & -0.78$\pm$0.12 \\
46 & 701 & -2138 & 2.74$\pm$0.15 & 22.42$\pm$0.13 & 7747$\pm$2 & 22$\pm$3 & 0.2$\pm$0.1 & -0.16$\pm$0.05 \\
\hline
A1 & 555 & 606 & 1.22$\pm$0.13 & 23.00$\pm$0.23 & 10491$\pm$4 &  &  & \\
A2 & 559 & 1020 & 1.56$\pm$0.15 & 23.17$\pm$0.12 & 10905$\pm$10 &  &  & \\
\hline
\hline
\end{tabular}

\footnotesize{$^*$The values of $r_{eff}$ and $\langle\mu_{eff}\rangle$ for G-20 in this
table are strongly discrepant with those presented in Chilingarian et al.
(2007c) measured on HST WFPC2, because here we did not take into account
seeing quality of our ground-based CFHT observations.}

\end{table*}

\begin{table*}
\caption{Selected Lick indices, values of derived values of [Mg/Fe]
using SSP models by Thomas et al. (2003) and brief comments concerning 
structures seen in colour maps.
\label{tablick}}
\begin{tabular}{lccccccl}
\hline
\hline
N & Fe$_{5015}$ & Mg$b$ & Fe$_{5270}$ & Fe$_{5335}$ & Fe$_{5406}$ & [Mg/Fe] & Comments \\
  &         \AA & \AA & \AA & \AA & \AA & dex & \\
\hline
01 & 3.89$\pm$0.28 & 3.40$\pm$0.10 & 2.59$\pm$0.10 & 2.05$\pm$0.10 & 1.78$\pm$0.09 & 0.22$\pm$0.10 & red core \\
02 & 4.61$\pm$0.19 & 3.07$\pm$0.07 & 2.46$\pm$0.07 & 2.45$\pm$0.09 & 1.70$\pm$0.07 & 0.14$\pm$0.07 & red core \\
03 & 3.07$\pm$0.54 & 2.70$\pm$0.21 & 2.17$\pm$0.21 & 1.28$\pm$0.21 & 0.64$\pm$0.20 & 0.20$\pm$0.20 \\
04 & 4.70$\pm$0.47 & 2.94$\pm$0.19 & 1.63$\pm$0.19 & 1.23$\pm$0.20 & 0.56$\pm$0.17 & 0.48$\pm$0.06 \\
05 & 3.96$\pm$0.35 & 2.78$\pm$0.14 & 3.01$\pm$0.14 & 2.17$\pm$0.14 & 1.41$\pm$0.11 & -0.11$\pm$0.15 & red core\\
06 & 5.46$\pm$0.12 & 3.67$\pm$0.05 & 2.93$\pm$0.05 & 2.50$\pm$0.05 & 1.38$\pm$0.04 & 0.11$\pm$0.04 & red core\\
07 & 2.64$\pm$0.19 & 1.39$\pm$0.07 & 1.75$\pm$0.07 & 1.12$\pm$0.09 & 1.21$\pm$0.06 & -0.02$\pm$0.09 & blue core \\
08 & 4.50$\pm$0.47 & 3.07$\pm$0.19 & 2.45$\pm$0.19 & 2.04$\pm$0.19 & 0.76$\pm$0.17 & 0.18$\pm$0.22 \\
09 & 2.85$\pm$0.28 & 3.21$\pm$0.11 & 2.99$\pm$0.11 & 1.69$\pm$0.11 & 1.40$\pm$0.09 & 0.08$\pm$0.12 \\
10 & 2.12$\pm$0.57 & 2.31$\pm$0.22 & 1.67$\pm$0.23 & 1.66$\pm$0.25 & 0.83$\pm$0.22 & 0.24$\pm$0.25 \\
11 & 4.73$\pm$0.32 & 2.64$\pm$0.14 & 2.12$\pm$0.12 & 2.00$\pm$0.12 & 1.69$\pm$0.11 & 0.15$\pm$0.17 & red core \\
12 & 2.01$\pm$0.56 & 2.29$\pm$0.22 & 0.95$\pm$0.24 & 2.61$\pm$0.24 & 0.57$\pm$0.22 & 0.41$\pm$0.27 \\
13 & 4.37$\pm$0.43 & 2.56$\pm$0.17 & 3.54$\pm$0.17 & 1.92$\pm$0.17 & 1.70$\pm$0.14 & -0.23$\pm$0.09 \\
14 & 3.58$\pm$0.31 & 2.41$\pm$0.12 & 2.34$\pm$0.12 & 0.96$\pm$0.13 & 0.82$\pm$0.10 & 0.17$\pm$0.16 \\
15 & 4.22$\pm$0.15 & 3.24$\pm$0.06 & 2.93$\pm$0.06 & 2.12$\pm$0.06 & 1.32$\pm$0.04 & 0.06$\pm$0.06 & red core\\
16 & 4.36$\pm$0.14 & 3.68$\pm$0.06 & 2.52$\pm$0.06 & 2.01$\pm$0.06 & 1.41$\pm$0.06 & 0.28$\pm$0.04 & red core\\
17 & 4.85$\pm$0.17 & 4.84$\pm$0.07 & 3.33$\pm$0.07 & 2.64$\pm$0.07 & 1.77$\pm$0.06 & 0.20$\pm$0.04 & red core\\
18 & 4.87$\pm$0.16 & 3.53$\pm$0.06 & 2.77$\pm$0.06 & 2.44$\pm$0.07 & 1.75$\pm$0.06 & 0.14$\pm$0.06 & red core\\
19 & 6.17$\pm$0.60 & 1.71$\pm$0.25 & 2.16$\pm$0.23 & 1.79$\pm$0.23 & 1.58$\pm$0.19 & -0.09$\pm$0.23 \\
20 & 5.57$\pm$0.16 & 4.79$\pm$0.06 & 3.25$\pm$0.06 & 2.96$\pm$0.06 & 1.95$\pm$0.06 & 0.18$\pm$0.04 \\
21 & 4.13$\pm$0.22 & 4.85$\pm$0.08 & 2.65$\pm$0.08 & 2.11$\pm$0.10 & 1.75$\pm$0.08 & 0.45$\pm$0.07 \\
22 & 4.26$\pm$0.41 & 3.06$\pm$0.15 & 3.37$\pm$0.15 & 1.82$\pm$0.17 & 1.21$\pm$0.14 & -0.10$\pm$0.17 \\
23 & 4.93$\pm$0.44 & 2.42$\pm$0.17 & 1.33$\pm$0.19 & 0.82$\pm$0.19 & 0.68$\pm$0.16 & 0.29$\pm$0.06 \\
24 & 4.55$\pm$0.13 & 5.01$\pm$0.05 & 2.94$\pm$0.05 & 2.53$\pm$0.05 & 1.72$\pm$0.04 & 0.34$\pm$0.03 & red core\\
25 & 4.56$\pm$0.20 & 3.53$\pm$0.07 & 2.80$\pm$0.07 & 2.19$\pm$0.08 & 1.42$\pm$0.07 & 0.14$\pm$0.07 & red core\\
26 & 3.95$\pm$0.57 & $...\pm...$   & 2.02$\pm$0.22 & 3.36$\pm$0.22 & 0.42$\pm$0.20 & $...\pm...$   \\
27 & 3.50$\pm$0.16 & 2.90$\pm$0.06 & 2.15$\pm$0.06 & 1.99$\pm$0.07 & 1.52$\pm$0.06 & 0.26$\pm$0.07 \\
28 & 4.63$\pm$0.21 & 3.62$\pm$0.08 & 2.42$\pm$0.08 & 1.86$\pm$0.08 & 1.28$\pm$0.07 & 0.32$\pm$0.06 & red core\\
29 & 2.59$\pm$0.49 & 2.91$\pm$0.18 & 2.23$\pm$0.19 & 2.35$\pm$0.19 & 3.16$\pm$0.15 & 0.16$\pm$0.24 \\
30 & 4.86$\pm$0.19 & 2.90$\pm$0.08 & 2.92$\pm$0.08 & 2.28$\pm$0.08 & 1.68$\pm$0.06 & -0.05$\pm$0.09 \\
31 & 3.56$\pm$0.15 & 3.50$\pm$0.05 & 2.61$\pm$0.05 & 2.13$\pm$0.05 & 1.31$\pm$0.05 & 0.21$\pm$0.05 & red core\\
32 & 3.74$\pm$0.19 & 3.55$\pm$0.07 & 2.70$\pm$0.07 & 2.38$\pm$0.07 & 1.71$\pm$0.06 & 0.15$\pm$0.07 & red core\\
33 & 3.81$\pm$0.21 & 3.03$\pm$0.08 & 2.50$\pm$0.08 & 2.06$\pm$0.09 & 1.53$\pm$0.08 & 0.15$\pm$0.09 \\
34 & 2.60$\pm$0.57 & 1.77$\pm$0.23 & 1.14$\pm$0.23 & 2.43$\pm$0.23 & 1.84$\pm$0.19 & 0.14$\pm$0.26 \\
35 & 1.54$\pm$0.24 & 1.97$\pm$0.09 & 2.80$\pm$0.09 & 1.65$\pm$0.10 & 0.93$\pm$0.09 & -0.25$\pm$0.10 & blue core\\
36 & 3.36$\pm$0.22 & 3.16$\pm$0.09 & 2.85$\pm$0.09 & 2.49$\pm$0.09 & 1.62$\pm$0.07 & 0.00$\pm$0.09 & red core\\
37 & 3.00$\pm$0.54 & 2.49$\pm$0.22 & 2.84$\pm$0.19 & 2.17$\pm$0.21 & 1.98$\pm$0.16 & -0.11$\pm$0.22 \\
38 & 1.14$\pm$0.09 & 1.55$\pm$0.04 & 1.42$\pm$0.04 & 1.28$\pm$0.04 & 0.77$\pm$0.03 & 0.12$\pm$0.04 & blue core\\
39 & 3.93$\pm$0.28 & 2.10$\pm$0.11 & 1.69$\pm$0.11 & 2.06$\pm$0.12 & 1.78$\pm$0.09 & 0.13$\pm$0.14 \\
40 & 3.92$\pm$0.13 & 4.24$\pm$0.05 & 2.84$\pm$0.05 & 2.34$\pm$0.05 & 1.60$\pm$0.04 & 0.25$\pm$0.04 & red core\\
41 & 3.35$\pm$0.16 & 3.90$\pm$0.07 & 2.82$\pm$0.07 & 2.33$\pm$0.07 & 1.60$\pm$0.05 & 0.19$\pm$0.05 & red core\\
42 & 3.31$\pm$0.69 & 3.40$\pm$0.25 & 1.51$\pm$0.27 & 1.61$\pm$0.27 & 1.40$\pm$0.22 & 0.43$\pm$0.06 \\
43 & 5.46$\pm$0.45 & 2.16$\pm$0.18 & 2.30$\pm$0.18 & 2.26$\pm$0.18 & 1.21$\pm$0.15 & -0.07$\pm$0.26 \\
44 & 4.79$\pm$0.18 & 3.80$\pm$0.07 & 2.67$\pm$0.07 & 2.27$\pm$0.07 & 1.55$\pm$0.06 & 0.22$\pm$0.06 & red core\\
45 & 2.56$\pm$0.43 & 3.07$\pm$0.16 & 1.89$\pm$0.16 & 1.90$\pm$0.16 & 0.48$\pm$0.15 & 0.34$\pm$0.15 \\
46 & -12.09$\pm$0.20 & 1.50$\pm$0.09 & 0.87$\pm$0.09 & 0.93$\pm$0.11 & 0.54$\pm$0.08 & 0.35$\pm$0.11 & strong star formation\\
\hline
\hline
\end{tabular}
\end{table*}

In this section, we present an atlas containing photometric and spectroscopic
information on 46 \object{Abell~496} members, and the tables with parameters of
morphology, integrated photometry, kinematics, and stellar populations for
these galaxies. For the two cluster members (marked as A1 and A2), where
stellar population fitting was not possible due to low signal-to-noise ratio
of the spectra, we provide redshifts and all photometrical and
morphological parameters.

We compute $B$ band absolute magnitudes in Table~\ref{tabmorph}, as well
as $B$ mean surface brightnesses in Table~\ref{tabfit} from $g'$
magnitudes using the photometrical transformation from Fukugita et al.
(1995), assuming the SED of elliptical galaxies for all objects ($B - g'
= 0.55$~mag, $R - r' = -0.25$~mag), also applying corrections for
Galactic extinction (Schlegel et al. 1998), cosmological dimming
(0.14~mag), and K-correction ($K_r' = 0.05$~mag, $K_g' = 0.1$~mag,
Fukugita et al. 1995).  Structural parameters: effective radii and
surface brightnesses have been calculated directly from the images,
without any bulge/disc decomposition or light profile fitting. Total
magnitudes in the four CFHT bands in Table~\ref{tabmorph} have been
corrected for Galactic extinction only.

The types of embedded structures in Table~\ref{tabmorph} are given as
follows: B, bar, D, disc, R, ring, S, spiral. Strengths of embedded
structures are specified as (s) for strong or (w) for weak.

For every galaxy we present:
\begin{enumerate}
\item{description of photometric, kinematical, and stellar population
properties;}
\item{cutout of the Megacam image in the $r'$ band with a size of 
23$\times$23~arcsec$^2$ for all galaxies for easy comparison of their spatial 
dimensions;}
\item{unsharp masking of the $r'$ band image, zoomed on the galaxy;}
\item{$u^* - r'$, $g' - r'$, and $g' - i'$ colour maps;}
\item{FLAMES/Giraffe spectrum, best-fitting PEGASE.HR model and residuals;}
\item{$\chi^2$ map in the age--metallicity space.}
\end{enumerate}

We applied an elliptically-smoothed unsharp masking technique (Lisker et
al. 2006) to the $r'$-band image using the ellipticity and orientation
of the isophotes of 23, 24, and 25 mag~arcsec$^{-2}$ in the $g'$ band.
We created masks for 8 different major axes of smoothing elliptical
Gaussians.  For every galaxy, we present a subjectively selected mask,
where embedded structures are most clearly seen. If no embedded
structures have been detected, the image, corresponding to the
orientation of the 25~mag~arcsec$^{-2}$ isophote and major axis
dispersion of 0.5~arcsec for the smoothing Gaussian, is shown.

To build colour maps, we have to take into account the differences between the
telescope PSF and atmosphere seeing for observations in different Megacam
bands.  Images in the $u^*$ and $g'$ bands have nearly the same seeing
quality, while $r'$ and $i'$ are slightly better. We have convolved the $r'$
and $i'$ band images with circular Gaussians having FWHM 0.57 and
0.34~arcsec to match the image quality in $u^*$ and $g'$.

To get reliable colour estimates at the peripheral parts of the
galaxies, where the signal level becomes rather low, we have applied
the Voronoi adaptive binning procedure (Cappellari \& Copin 2003) to
the images in $r'$ to get a constant signal-to-noise ratio of
50 per bin. All tessellae containing more than 120 pixels are
masked on the presented colour maps.

\subsection{Notes on individual objects}

{\bf ACO496J043306.97-131238.8 (G-01)}: This early-type
spiral exhibits an extended disc and a small bar, which is aligned
nearly along the minor axis of the external disc, explaining the
isophotal twisting. The colour maps indicate the presence of a small
red bulge and the velocity dispersion is in agreement with a 
low-mass bulge. Intermediate age and slightly sub-solar metallicity
are normal for such a low-luminosity spiral.  However, [Mg/Fe] is
super-solar and is among the highest values obtained for such faint
galaxies.  The galaxy is located in the outer parts of the bright
X-ray halo observed with Chandra (Dupke \& White 2003) and XMM-Newton
(Tanaka et al. 2006). Its projected distance (d$_{proj}$ = 304~kpc) places
it far from the cluster core.  No emission lines are seen
in the spectrum.

{\bf ACO496J043308.85-130235.6 (G-02)}: This late type spiral is nearly
edge-on. A dust lane is seen on the westside of the galaxy on the colour
maps.
Star formation regions in the southern part of the disc seem strongly
obscured by dust. The residuals of the fitting exhibit a narrow H$\beta$
line in emission, indicating that HII regions are found in the
centre. As for the previous galaxy, the stellar population is not older
than 4~Gyr, and the metallicity is slightly sub-solar. The velocity
dispersion is normal for a galaxy, which is among the brightest objects
of our sample, but clearly not an early-type galaxy. The [Mg/Fe]
abundance ratio is somewhat higher than the solar value.  No X-ray
emission is found at the location of ACO496J043308.85-130235.6 (G-02)
and its projected distance (d$_{proj}$ = 557~kpc) places it at more than
one third of the cluster virial radius. Our radial velocity for
ACO496J043308.85-130235.6 (10862~km~s$^{-1}$) is strongly discrepant
with the one reported by NED (10197~km~s$^{-1}$), but two other studies
(Malumuth et al. 1992; Christlein \& Zabludoff 2003) give heliocentric
radial velocities different by less than 40~km~s$^{-1}$ from our value.

{\bf ACO496J043312.79-130449.3 (G-03)}: This faint and
regular dwarf elliptical galaxy is not nucleated and has no colour
gradient.  Its low velocity dispersion, relatively old age, very
metal-poor stellar population, and low surface brightness make it a
normal dwarf elliptical.  
Its projected distance (d$_{proj}$ = 470~kpc) is large and the galaxy
is found in the outskirts of the X-ray halo. No sign of ram-pressure
stripping or tidal interaction is observed, as expected if the galaxy
is on a roughly circular orbit or just entering the X-ray halo.

{\bf ACO496J043317.75-131536.6 (G-04)}: This dwarf
lenticular or elliptical galaxy has a 
smooth and elongated appearance. The unsharp masked image seems to
indicate the presence of a faint bar or nearly edge-on internal disc,
but this embedded structure is certainly faint since it is not seen on
colour maps.  With a low velocity dispersion, an intermediate age for
its stellar population and a correspondingly low metallicity, this
galaxy does not differ strongly from ``classical'' dwarfs.
Both its projected distance (d$_{proj}$ = 183~kpc) and its radial
velocity suggest that interaction with the hot intracluster medium
could have induced a short star formation episode in the
ACO496J043317.75-131536.6 (G-04) centre. Despite a low S/N spectrum, we can
be sure that no emission lines are present. The radial velocity reported
by NED (Christlein and Zabludoff, 2003) differs by more than
300~km~s$^{-1}$ from our value.

{\bf ACO496J043318.95-131726.9 (G-05)}: The central part of this dwarf
elliptical has redder colours than the outskirts and the velocity
dispersion corresponds to the mean value for this absolute
magnitude. However the age of the stellar population is close to the
maximum value allowed with our PEGASE.HR database (17.6~Gyr) both with
quite a low metallicity. Like for the previous galaxy, its projected
distance (d$_{proj}$ = 184~kpc) and its radial velocity suggest
ACO496J043318.95-131726.9 (G-05) is on a trajectory crossing the cluster
core. But the progenitor of this dwarf could have already consumed most
of its gas when it entered the X-ray halo and part of its disc could
have been tidally stripped, leaving a low-mass bulge nearly intact.

{\bf ACO496J043320.35-130314.9 (G-06)}: This lenticular
galaxy has an intermediate luminosity and is the brightest galaxy of
the spectroscopic sample. The unsharp masking technique reveals a
strong bar and ring system, which is also redder on the colour maps
compared to the underlying disc.  The galaxy has an intermediate age,
a slightly metal poor and $\alpha$-overabundant stellar population,
with a quite normal velocity dispersion for the bulge of an S0
galaxy. No emission lines are detected in the spectrum.
ACO496J043320.35-130314.9 (G-06) is found outside the X-ray halo in the
northern part of the cluster.

{\bf ACO496J043321.37-130416.6 (G-07)}: This dwarf lenticular or
early-type spiral is very peculiar since the elongated central structure
is bluer than the underlying disc. Both the unsharp masked image and
colour maps exhibit a spiral pattern extending over most of the inclined
disc. An [OIII] line is noticed in emission on the residual after
fitting the stellar population but H$\beta$ is located outside the
observed wavelength range. This galaxy has the second bluest ($u^* -
r'$) colour index in an aperture of 1.2~arcsec of our whole sample.  The
stellar population fitting gives a very young age and a slightly
sub-solar metallicity, but both estimations have 
large uncertainties and we suspect that a stellar population with an age
of 1.2~Gyr and a metallicity of [Fe/H]=-0.55~dex could also be present
in the galaxy centre, indicating that star formation began at that time
and is continuing up to now. The [Mg/Fe] ratio is nearly solar and the
low velocity dispersion is determined with good precision.  Our value
for the radial velocity (7736~km~s$^{-1}$) differs by 78km~s$^{-1}$ from
the value given NED and found by Durret at al. (1999), but this
difference is only equal to 1.2$\sigma$ of the older measurement.  As
for the previous galaxy, ACO496J043321.37-130416.6 (G-07) is located at
the northern border of the X-ray halo, but its radial velocity is
smaller than the mean cluster velocity by more than
2000~km~s$^{-1}$. This dwarf could be orbiting around the more luminous
lenticular ACO496J043320.35-130314.9 (G-06), but their very high
velocity difference and their separation of about 1~arcmin (40~kpc)
indicate that if a strong tidal interaction has occurred in the past,
leading to a burst of star formation in the centre, the tidal forces are
not acting anymore.

{\bf ACO496J043324.61-131111.9 (G-08)}: This is a faint regular dE
galaxy, which could be nucleated as evidenced by a slightly bluer centre
in the colour maps. The faint S/N ratio of the spectrum does not allow
us to determine the age and metallicity precisely, but the stellar
population seems to have an 
old age and low metallicity, both consistent with the values generally
expected in such dwarfs. Only the central velocity dispersion can be
determined with a good precision since the absorption lines are clearly
very narrow. Its velocity is smaller than the mean cluster velocity and
its projected distance to the cluster centre is 209~kpc, which is about
half the radius of the X-ray halo.

{\bf ACO496J043324.91-131342.6 (G-09)}: This dE exhibits a slightly
concentrated structure revealed by the unsharp masking technique and a
colour gradient with a redder colour in the centre. The outer part has
colour indices typical of those seen in the discs of lenticulars in our
sample. Its velocity dispersion is low, and the age of its stellar
population is quite old, with a low metallicity and a nearly solar abundance
ratio. No emission lines are detected.  Its velocity is about
700~km~s$^{-1}$ lower than the mean cluster velocity. The projected
distance to the cluster centre is 139~kpc; at about 40~kpc in projection, a
brighter elliptical (M$_B\simeq -20$ and
$\sigma$=207~km~s$^{-1}$) is found with a radial velocity of
9300~km~s$^{-1}$, giving an argument that ACO496J043324.91-131342.6
(G-09) could be undergoing tidal interaction.

{\bf ACO496J043325.10-130906.6 (G-10)}: This dE is faint
and does not show any substructures. Only a slight colour gradient
appears on colour maps, but there is no signature of a nucleus.  Stellar
population parameters are uncertain because of low signal-to-noise
ratio in the spectrum.
No emission lines are detected. The velocity dispersion is 
low, but higher than for galaxies of similar luminosity. This faint
dwarf is in projection inside the X-ray halo at d$_{proj}$ = 273~kpc.

{\bf ACO496J043325.40-131414.6 (G-11)}: This is a
faint dE galaxy with no substructure, except a slightly concentrated
red central part. This red centre is, however, only marginally resolved
and could be considered as a nucleus. The galaxy exhibits an
intermediate age, a metal poor stellar population and a quite normal
velocity dispersion for such a faint dwarf. No emission lines are
detected in the spectrum. Like the previous galaxy, this dwarf is in
projection near the cluster core (d$_{proj}$ = 130~kpc), but also very
near another dE, ACO496J043324.91-131342.6 (G-09).  However their radial
velocities differ by 825~km~s$^{-1}$, putting them in different
physical areas in the cluster. 

{\bf ACO496J043325.54-130408.0 (G-12)}: This low surface brightness dE
is quite regular and with a small gradient in colour changing from an
intermediate colour in the centre to a blue one in the outer
regions. The poor S/N ratio spectrum still allows us to determine
metal-poor intermediate age stellar population.  The velocity dispersion
is low, as expected for its faint luminosity. No emission lines are
present in the residual of the fitting.
The projected distance to the cluster centre is 449~kpc and the galaxy
is located just at the northern border of the X-ray halo.

{\bf ACO496J043326.49-131717.8 (G-13)}: This dwarf lenticular exhibits a
faint elongated structure.  This inclined disc may contain a spiral pattern,
which appears bluer on some colour maps despite the noise.  In the galaxy
centre a red, compact, and slightly elongated component could be a small
bulge.  Its velocity dispersion is quite low, with an intermediate age and a
low metallicity. This object has the lowest value of [Mg/Fe] of the sample,
indicating that subsequent star formation episodes might have occurred in
this object. No emission lines are detected.
It is located quite close to the cluster centre, at a projected
distance of 120~kpc.

{\bf ACO496J043329.79-130851.7 (G-14)}: This is a faint dE galaxy with
no substructure except for a trend to be centrally concentrated. A
strong colour gradient is observed, with redder colours in the centre
and becoming bluer toward the outskirts. Its velocity dispersion is
low. One can notice two minima of $\chi^2$ in the age-metallicity
space. This can be considered as a hint for a complex star formation
history in this object, however, higher signal-to-noise ratio in the
spectrum is required to disentangle the two star formation episodes.
No emission lines are detected. 
The projected distance to the cluster centre is 267~kpc.

{\bf ACO496J043331.48-131654.6 (G-15)}: This low-luminosity lenticular
galaxy exhibits prominent Z-shaped structure, revealed by the unsharp
masking technique, which is probably a spiral pattern in the strongly
inclined stellar disc. The stellar population of the galaxy is
moderately old, with sub-solar metallicity and exactly solar [Mg/Fe]
ratio. No emission lines are detected. The velocity dispersion is 
low compared to other galaxies of similar luminosities, suggesting that
ACO496J043331.48-131654.6 (G-15) may be a stripped spiral or a
later-type galaxy with a small bulge (see colour maps). The galaxy is
located only 75~kpc in projected distance from the cluster cD, and its
radial velocity is close to the cluster redshift. However, its
photometric, dynamical, and stellar population properties resemble other
galaxies in our sample located much further away from the cluster
centre, so this is an indirect argument that its real distance from cD
could be considerably larger than the projected one.
ACO496J043333.17-131712.6 (G-17), located at the same projected distance
and only 30~arcsec away from ACO496J043331.48-131654.6 (G-15) has a very
similar luminosity, but completely different properties. Our radial
velocity value for ACO496J043331.48-131654.6 (G-15), 9586~km~s$^{-1}$,
is strongly discrepant with the one reported by NED, 9875~km~s$^{-1}$,
but agrees with the HyperLEDA value (Malumuth et al. 1992).  However,
the NED value is very close to the redshift of ACO496J043333.17-131712.6
(G-17), so we conclude that it might be a cross-identification mistake.

{\bf ACO496J043332.07-131518.1 (G-16)}: This low-luminosity elliptical
galaxy is found in the inner part of the cluster (d$_{proj}$ = 54~kpc). It
shows a red concentrated structure in the centre.  The stellar population is
quite old and metal-poor, and [Mg/Fe] is super-solar. 
ACO496J043332.07-131518.1 (G-16) exhibits a moderate velocity dispersion
value and no emission lines. Its radial velocity does not differ from the
mean cluster redshift. The properties of its central stellar population are
very unusual for elliptical galaxies and even for low-luminosity objects;
they are very similar to those of ACO496J043346.71-131756.2 (G-31), an S0
among the most luminous in our spectroscopic sample. If the disc of a
galaxy, with a bulge similar to the one of the previous S0, has lost most of
its stars at an early phase of its evolution during tidal interactions, it
is plausible that the result will look like this low-luminosity E, a spheroid
with a colour gradient. An alternative hypothesis is that the gas has been
swept very rapidly after the galaxy formation, leaving it quiescent for a
long time in the cluster core without any further merger with other objects.

{\bf ACO496J043333.17-131712.6 (G-17)}: This
low-luminosity elliptical galaxy belongs to the population of 
unusual galaxies in the central dense part of the Abell 496 cluster
(72~kpc in projected distance). Its stellar population of high
velocity dispersion, very old age, metal-rich and $\alpha$-enhanced,
and 
compact appearance make it very different from nearby E
galaxies of this luminosity. No emission lines are detected. No presence of
embedded structures is revealed by the unsharp masking. The central
part of the galaxy is relatively red and extended; the stellar
population properties correspond to those of a normal bulge as seen in
more massive lenticulars.  Its radial velocity is very close to the
mean cluster velocity.

{\bf ACO496J043333.53-131852.6 (G-18)}: This is an
early-type galaxy that can be classified as a low-luminosity SB0a
with an absolute magnitude that places it slightly above the brightest
dE/dS0.  Unsharp masking reveals a prominent bar aligned nearly along
the minor axis of the external isophotes. The difference in colour
between the bulge and disc is clearly seen on all the colour maps, as
well as for the bar which is also redder compared to the disc.  Its
intermediate age, nearly solar metallicity, and moderate velocity
dispersion resemble properties of objects of this luminosity in the
nearby Universe.  However, the [Mg/Fe] abundance ratio is somewhat
higher than solar, indicating a 
short period of star formation ($<$1.5~Gyr).  No emission lines are
detected.  The galaxy is located in the central part of the bright X-ray
halo at a projected distance of 130~kpc.

{\bf ACO496J043334.54-131137.1 (G-19)}: This is a
faint dE galaxy showing no evidence for embedded structures but a
central component redder than the outer parts, however, not as red  as
typical bulges already seen in the other brighter galaxies.  Its
velocity dispersion is low, and its stellar population has an intermediate
age and significantly sub-solar metallicity. No emission lines are detected.  
This galaxy is located quite close to the cluster centre, at a projected
distance of 156~kpc.

{\bf ACO496J043337.35-131520.2 (G-20)}: This new M32-like
compact elliptical has been extensively discussed by Chilingarian et
al. (2007c) and is located in projection on the central cD halo.  The
central very bright component of this very peculiar dwarf has colours
similar to most of the other ellipticals and lenticulars. With its
high velocity dispersion for its luminosity, very old stellar
population, nearly solar metallicity and mild $\alpha$-enhancement, it
seems that ACO496J043337.35-131520.2 (G-20) is the disturbed remnant of an
intermediate luminosity S0 where most of the disc has been tidally
stripped.

{\bf ACO496J043338.22-131500.7 (G-21)}: This dwarf elliptical exhibits
very different structural properties from the previous galaxy (A496cE),
although it is located only at a projected distance of about 30~kpc of
the cD (Chilingarian et al., 2007b). On the other hand, the colours in
the central parts of the galaxy seem 
quite similar in both objects, and the stellar population in
ACO496J043338.22-131500.7 (G-21) is only slightly more metal poor than
that of ACO496J043337.35-131520.2 (G-20) and with higher
$\alpha$-enhancement, while their ages are nearly the same.  The
velocity dispersion of ACO496J043338.22-131500.7 (G-21) is even larger
and more unusual for such a small object. If the tidal stripping of a
larger disc galaxy could also be invoked to explain the spectral
properties, in the present case the remnant structural properties are
quite normal compared to bulges. Two objects are reported in NED near
the position of this dwarf and for the second one the radial velocity
(Quintana \& Ramirez, 1990) is 400~km~s$^{-1}$ smaller than our value.

{\bf ACO496J043339.07-131319.7 (G-22)}: This low-luminosity dwarf
elliptical may host a faint nucleus as evidenced by the unsharp mask
image. The colour maps are noisy, but indicate some colour difference
between the centre and the outer parts. A low velocity dispersion, a
relatively old and 
metal-poor stellar population, and low surface
brightness make it a 
normal dwarf elliptical. No emission lines
are detected. Its radial velocity is larger by 900~km~s$^{-1}$ than the
mean cluster velocity; it is projected quite close to the cluster
centre, at a projected distance of 89~kpc, but its real distance from
the centre could be larger.

{\bf ACO496J043339.72-131424.6 (G-23)}: The centre of
this faint dE is blue and could be slightly bluer than the rest of the
galaxy despite the noise in the colour maps.  Its velocity
dispersion is low, and the age of its stellar population is 
old and metal-poor.
No emission lines are detected.  
The projected distance to the cluster centre is 51~kpc, but there is no
obvious reason to put this dE into the cluster core.

{\bf ACO496J043341.69-131551.8 (G-24)}: This very unusual,
low-luminosity elliptical, representative of the peculiar galaxy
population in the inner part of the cluster, is projected on the outer
parts of the cD halo (d$_{proj}$ = 36~kpc). This galaxy is the highest
velocity dispersion object in our sample. Its velocity dispersion and
stellar population with slightly sub-solar metallicity, high
$\alpha$-enhancement, and very old age are reminiscent of bulges of
bright lenticular galaxies with a luminosity.  This galaxy is probably a
``scaled-up'' analogue of A496cE (ACO496J043337.35-131520.2 (G-20);
Chilingarian et al., 2007b) and the evolutionary scenarii of these two
objects are very similar: heavy tidal stripping of a more luminous S0.

{\bf ACO496J043342.10-131653.7 (G-25)}: This low-luminosity lenticular
galaxy is found in the inner part of the cluster (d$_{proj}$ =
60~kpc). Unsharp masking as well as colour maps reveal a bar or ringlike
structure and faint spiral arms in the stellar disc.  The stellar
population is quite old and metal-poor, although [Mg/Fe] is super-solar,
which is rather uncommon for such a faint S0 galaxy, and its velocity
dispersion is average.  No emission lines are detected.

{\bf ACO496J043342.13-131258.8 (G-26)}: This galaxy has
the second largest effective radius (2.8~kpc) among the sample of
early-type objects, while its luminosity is moderate, making it a low
surface brightness galaxy. The projected distance to the centre is
d$_{proj}$ = 109~kpc and the radial velocity is higher by
465~km~s$^{-1}$ than the mean value for the cluster, but this object
does not resemble other galaxies populating the central region of the
cluster. The velocity dispersion is 
low and the stellar population properties are normal for its
luminosity. There was an artifact in the spectrum around 5320~\AA\
(5140~\AA\ in rest frame) so the Mg$b$ Lick index could not be measured.

{\bf ACO496J043342.83-130846.8 (G-27)}: This is a quite
low-luminosity S0/E galaxy with an object superimposed less than
2~arcsec from its centre but otherwise no obvious substructures. No
colour gradient is observed in this galaxy. Its velocity dispersion is
quite low, with a medium age of its stellar population, a quite low
metallicity and a high value of [Mg/Fe].  No emission lines are
detected. 
The projected distance to the cluster centre is 264~kpc and
the galaxy is located in the outer parts of the X-ray halo.

{\bf ACO496J043343.04-130514.1 (G-28)}: This is a 
bright dS0 galaxy with no substructure. A strong colour gradient is
observed: the very central region is redder than the rest of the
galaxy and could be a small bulge (marginally resolved in direction of
the minor axis). Its velocity dispersion is quite low, and the age of
its stellar population is 
old, with a low metallicity but a high value of [Mg/Fe].  No emission
lines are detected.
The projected distance to the cluster centre is 396~kpc. The galaxy,
located at the border of the X-ray halo, seems to belong to a group of
galaxies for which 4 members are in the spectroscopic sample, with
ACO496J043349.08-130520.5 (G-33) being the brightest and largest
galaxy.

{\bf ACO496J043343.04-125924.4 (G-29)}: This inclined or elongated faint
dS0/dE, with no evident embedded substructure, exhibits a weak colour
gradient with redder colours in the centre and becoming bluer towards
the outskirts. Its velocity dispersion is low, and the age of its
stellar population is quite old, with a very low metallicity.  No
emission lines are detected.
The projected distance to the cluster centre is 614~kpc and the galaxy
is found in the north, away from the X-ray halo.

{\bf ACO496J043345.67-130542.2 (G-30)}: This moderately
bright dS0 galaxy with no substructures belongs to the group mentioned
in the description of ACO496J043343.04-130514.1 (G-28). A colour gradient is
observed, with the central region redder than the rest of the galaxy,
indicating a small bulge. Its velocity dispersion is small and the age
of its stellar population is young, with a low metallicity and [Mg/Fe]
ratio.  No emission lines are detected. The projected distance to the
cluster centre is 382~kpc. Its velocity of 9689~km~s$^{-1}$ is
200~km~s$^{-1}$ smaller than the mean cluster velocity. Our
measurement of the radial velocity solves the problem of the
discrepancy between the values of 9770 $\pm$ 80 ~km~s$^{-1}$ and 9528
$\pm$ 101 ~km~s$^{-1}$ found respectively by Christlein \& Zabludoff
(2003) and Malumuth et al. (1992).

{\bf ACO496J043346.71-131756.2 (G-31)}: This is a
low-luminosity, early-type galaxy slightly brighter than
ACO496J043333.53-131852.6 (G-18). Faint spiral structure and a bar aligned
along the major axis of the galaxy are revealed by means of unsharp
masking. The galaxy exhibits a very old, metal poor,
$\alpha$-overabundant stellar population with a velocity dispersion
quite normal for its luminosity. No emission lines are detected. 
The galaxy is located some 120~kpc away from the cD galaxy
in projection, its radial velocity differs significantly
(-1500~km~s$^{-1}$) from the mean cluster redshift.

{\bf ACO496J043348.59-130558.3 (G-32)}: This quite faint
dE galaxy with no embedded substructure is the third member of the
group discovered at the northern border of the X-ray halo. As for the
other lenticulars of this group, a small red and compact central
structure is seen in the colour maps. Its velocity dispersion is quite
low and the age of its stellar population is very old and metal poor
but with a 
high [Mg/Fe] ratio.  No emission lines are detected.
The projected distance to the cluster centre is 402~kpc.

{\bf ACO496J043349.08-130520.5 (G-33)}: This early-type
galaxy can be classified as a dS0. Its luminosity places it among the
brightest representatives of the dE/dS0 class. Unsharp masking with
elliptical blurring corresponding to the shape of the $\mu _{B}$ = 24
mag/arcsec$^{-2}$ isophote reveals faint spiral arms, reminiscent of
structures observed in brighter dE/dS0 galaxies in the Virgo (Jerjen
et al. 2000; Barazza et al. 2002; Lisker et al. 2006) and Coma (Graham
et al. 2003) clusters. These spiral arms cause an isophote twist in
the outer regions of the galaxy.  We do not observe emission lines in
the spectrum of ACO496J043349.08-130520.5 (G-33). The old age and
low metallicity of this galaxy correspond well to similar galaxies in
the Virgo cluster. The velocity dispersion value is exactly what is
expected from an object of this luminosity. The [Mg/Fe] ratio is
slightly supersolar.  This galaxy is located some 0.4~Mpc in projection
from the centre of the cluster, being the brightest member of the
association including three other dwarf galaxies
(ACO496J043348.59-130558.3 (G-32); ACO496J043343.04 -130514.1 (G-28);
ACO496J043345.67-130542.2 (G-30)) located 24, 55, and 34 kpc in
projected distance and having very similar radial velocities, velocity
dispersions and stellar populations.

{\bf ACO496J043350.17-125945.4 (G-34)}: This is a faint, 
blue dS0 galaxy with no substructure and no obvious colour
gradient.  Its velocity dispersion is low, and the age of its stellar
population is young, with a very low metallicity.  
No emission lines are detected.  
The projected distance to the cluster centre is 610~kpc and the galaxy
is located in the north, well outside the X-ray halo.

{\bf ACO496J043351.54-131135.5 (G-35)}: This is a faint
dS0/dE galaxy with a very blue centre. This central compact structure
appears notably bluer than the outer zones and is surrounded by an
area with a colour gradient going from redder to bluer colours with
increasing radius.  Its velocity dispersion is small, and its stellar
population is very young, and metal-poor. 
No emission lines are detected.  
The projected distance to the cluster centre is 199~kpc and this
galaxy is located at more than half the X-ray halo radius.

{\bf ACO496J043352.77-131523.8 (G-36)}: As for the dS0
ACO496J043349.08-130520.5 (G-33), the unsharp masking with elliptical
blurring corresponding to the shape of the $\mu _{B}$ = 24
mag/arcsec$^{-2}$ isophote reveals a spiral pattern with very faint and
slightly blue spiral arms in this bright dS0. These spiral arms cause an
isophote twist in the outer regions of the galaxy. The spectral
properties of the stellar population are also very similar to those of
ACO496J043349.08-130520.5 (G-33), only differing in the [Mg/Fe] ratio,
which is nearly solar. The velocity dispersion is in exactly the same
range and corresponds to what is expected for a galaxy of this
luminosity. In conclusion, one can suspect that this dS0 is projected
just onto the cluster core (d$_{proj}$ = 137~kpc), since the other dS0
similar to this one is located outside the X-ray halo.

{\bf ACO496J043355.55-131024.9 (G-37)}: This is a very
faint dS0/dE galaxy with no obvious substructures.  A weak colour
gradient is observed, with colours redder in the centre and becoming
bluer towards the outskirts.  Its velocity dispersion is small, and
its stellar population is 
old having significantly sub-solar metallicity. No emission lines are
detected. It seems this dwarf has a similar stellar population in the
centre as ACO496J043318.95-131726.9 (G-05), but with a bulge mass four
times smaller.
The projected distance to the cluster centre is 256~kpc and the galaxy
is located in the external parts of the X-ray halo.

{\bf ACO496J043356.18-125913.1 (G-38)}: 
This is a bright dE/dS0 galaxy with an early-type morphology and
quite unusual stellar population properties.
A very blue core is seen
on colour maps, partly surrounded by a redder zone, and red patches
further out aligned along the major axis. Its velocity dispersion is
small, and the age of its stellar population is extremely young, with
a sub-solar metallicity.  Strong
H$\beta$ and [OIII] emission lines are detected, indicating that HII
regions are also found in the centre, implying an ongoing star
formation process. No signature of [NI] ($\lambda=5199$~\AA) is 
seen, which excludes the possibility of the shockwave gas ionisation.
The projected distance to the cluster centre is large (642~kpc) so
the burst of star formation is probably not due to the interaction 
with the intracluster medium.

{\bf ACO496J043359.03-130626.7 (G-39)}: This is a 
bright extended dS0 galaxy with a low surface brightness and a faint
compact central structure. Only a weak colour gradient is observed, with
the central region not as blue as the rest of the disc. Its velocity
dispersion is low, and the age of its stellar population is quite young,
with a low metallicity and a [Mg/Fe] ratio close to solar No emission
lines are detected.
The projected distance to the cluster centre is 398~kpc with the
galaxy location at the border of the X-ray halo.

{\bf ACO496J043401.57-131359.7 (G-40)}: Although this
galaxy is not in the cluster core, its spectroscopic and photometrical
properties are very similar to the ones of the peculiar ellipticals
found in the vicinity of the cD. This bright S0/E has intermediate
value of absolute magnitude and a velocity dispersion between those of
the two bright Es with the largest sigma values at less than 100~kpc
of the cD, ACO496J043341.69-131551.8 (G-24) and ACO496J043333.17-131712.6 (G-17). As
for these two galaxies, the stellar population with low metallicity
and $\alpha$-enhancement and very old age are similar to bulges of
brighter lenticular galaxies. Its 
projected distance to the cluster centre is 227~kpc. Since no massive
object is seen near this peculiar early-type galaxy in projection, one
can suspect that a radial orbit almost in the plane of the sky could
have brought it near the cD, where strong tidal effects have been
responsible for the tidal stripping of the disc and shrinking of the
bulge remnant.

{\bf ACO496J043403.19-131310.6 (G-41)}: With almost the
same luminosity as ACO496J043320.35-130314.9 (G-06), this SB0/SBa galaxy
exhibits an extended spiral pattern and a small bulge evidenced by its
red colour on the colour maps. Unsharp masking reveals a bar or
ringlike structure. Unexpected for a low-luminosity early-type
spiral, it has an old age, but the rest of its stellar population
properties are not unusual: it is metal poor and
$\alpha$-overabundant, with a quite normal velocity dispersion for a
low-mass bulge.  No emission lines are detected in the spectrum.
ACO496J043403.19-131310.6 (G-41) is found in the outer eastern part of the
X-ray halo.

{\bf ACO496J043408.50-131152.7 (G-42)}: This is a faint
dS0/dE galaxy with an elongated structure that could be a bar or an
inclined disc. A weak colour gradient is observed, with  redder colours
in the centre and becoming bluer towards the outskirts.  Its velocity
dispersion is small, and the age of its stellar population is 
old, with a very low metallicity. 
No emission lines are detected. 
The projected distance to the cluster centre is 315~kpc and the galaxy
is located just at the limit of the X-ray halo.

{\bf ACO496J043410.60-130756.7 (G-43)}: This is a 
faint dS0 galaxy with no substructure. A strong colour gradient is
observed, with redder colours in the centre and becoming bluer towards
the outskirts. Its velocity dispersion is low, and the age of its
stellar population is intermediate, with a correspondingly low
metallicity and a nearly solar [Mg/Fe] ratio.  No emission lines are
detected. 
The projected distance to the cluster centre is 418~kpc. This dS0
seems isolated and is located outside the X-ray halo.

{\bf ACO496J043413.00-131003.5 (G-44)}: This is a rather
bright face-on S0 or elliptical galaxy with no obvious substructures,
but very extended and with a low surface brightness except in the
innermost 2~arcsec. A colour gradient is observed: the very centre of
the galaxy appears redder than the outer zones. Its velocity
dispersion is average, and its stellar population is old, with a low
metallicity and a high [Mg/Fe] ratio.  No emission lines are
detected. 
The projected distance to the cluster centre is 385~kpc and it is
located just outside the X-ray halo limit.

{\bf ACO496J043413.08-131231.6 (G-45)}: This is a faint dE galaxy with
no substructure except for a trend to be centrally concentrated. A
strong colour gradient is observed, with redder colours in the centre
and becoming bluer towards the outskirts. Its velocity dispersion is
rather normal for its luminosity, and the age of its stellar population
is quite old, with a very low metallicity.  No emission lines are
detected.
The projected distance to the cluster centre is 344~kpc. This faint dE
is similar to another dwarf with the same properties,
ACO496J043329.79-130851.7 (G-14). Both galaxies do not seem to be
suffering an interaction with the intracluster medium or tidal
effects.

{\bf ACO496J043415.37-130823.5 (G-46)}: is a late type
star forming dwarf galaxy with a radial velocity differing from that
of the cluster by $-2141$~km~s$^{-1}$. On the unsharp masked image and
colour maps one can see several spatially unresolved star forming
regions. The spectrum shows strong [OIII] emission lines (H$\beta$ is
not in our wavelength range) and no obvious sign for emission in
[NI]$\lambda$5199, implying that this is a typical HII region
spectrum. The velocity dispersion is among the lowest in the sample.
The projected distance to the cluster centre is 440~kpc.

{\bf ACO496J043411.72-131130.2 (G-A1)}: This galaxy is the
faintest in our sample, and the smallest except for the cE galaxy.
The projected distance to the cluster centre is 348~kpc.
The spectrum was too faint to fit a stellar population model.

{\bf ACO496J043414.54-131303.0 (G-A2)}: This galaxy is the second
faintest in our sample.  The projected distance to the cluster centre
is slightly larger than 350 kpc and its spectrum was too faint
to fit a stellar population model.

\clearpage
\begin{figure*}
\caption{{\bf ACO496J043306.97-131238.8} (G-01). (a): CFHT Megacam $r'$-band image;
(b): result of unsharp masking; (c)-(e): adaptively binned colour maps;
(f) spectrum (black) and $\pm1\sigma$ flux uncertainties (red), 
best-fitting PEGASE.HR model (blue) and residuals of the fit; (g) map of the
$\chi^2$ on a wide grid of age and metallicities. Rectangle on the (a) panel
shows the dimensions of unsharp masking and colour maps. Contours on (b)-(e)
correspond to the surface brightness in $r'$, the outer contour is 
25~mag~arcsec$^{-2}$, the step is 1~mag~arcsec$^{-2}$.
\label{fig143677}}
\end{figure*}

\begin{figure*}
\caption{{\bf ACO496J043308.85-130235.6} (G-02). The same as in
Fig.~\ref{fig143677}
\label{fig160522}}
\end{figure*}

\clearpage
\begin{figure*}
\caption{{\bf ACO496J043312.08-130449.3} (G-03). The same as in
Fig.~\ref{fig143677}
\label{fig157963}}
\end{figure*}

\begin{figure*}
\caption{{\bf ACO496J043317.75-131536.6} (G-04). The same as in
Fig.~\ref{fig143677}
\label{fig138549}}
\end{figure*}

\clearpage
\begin{figure*}
\caption{{\bf ACO496J043318.95-131726.9} (G-05). The same as in
Fig.~\ref{fig143677}
\label{fig136223}}
\end{figure*}

\begin{figure*}
\caption{{\bf ACO496J043320.35-130314.9} (G-06). The same as in
Fig.~\ref{fig143677}
\label{fig159652}}
\end{figure*}

\clearpage
\begin{figure*}
\caption{{\bf ACO496J043321.37-130416.6} (G-07). The same as in
Fig.~\ref{fig143677}
\label{fig158915}}
\end{figure*}

\begin{figure*}
\caption{{\bf ACO496J043324.61-131111.9} (G-08). The same as in
Fig.~\ref{fig143677}
\label{fig147347}}
\end{figure*}

\clearpage
\begin{figure*}
\caption{{\bf ACO496J043324.91-131342.6} (G-09). The same as in
Fig.~\ref{fig143677}
\label{fig142386}}
\end{figure*}

\begin{figure*}
\caption{{\bf ACO496J043325.10-130906.6} (G-10). The same as in
Fig.~\ref{fig143677}
\label{fig150275}}
\end{figure*}

\clearpage
\begin{figure*}
\caption{{\bf ACO496J043325.40-131414.6} (G-11). The same as in
Fig.~\ref{fig143677}
\label{fig141659}}
\end{figure*}

\begin{figure*}
\caption{{\bf ACO496J043325.54-130408.0} (G-12). The same as in
Fig.~\ref{fig143677}
\label{fig159183}}
\end{figure*}

\clearpage
\begin{figure*}
\caption{{\bf ACO496J043326.49-131717.8} (G-13). The same as in
Fig.~\ref{fig143677}
\label{fig135128}}
\end{figure*}

\begin{figure*}
\caption{{\bf ACO496J043329.79-130851.7} (G-14). The same as in
Fig.~\ref{fig143677}
\label{fig053255}}
\end{figure*}

\clearpage
\begin{figure*}
\caption{{\bf ACO496J043331.48-131654.6} (G-15). The same as in
Fig.~\ref{fig143677}
\label{fig037015}}
\end{figure*}

\begin{figure*}
\caption{{\bf ACO496J043332.07-131518.1} (G-16). The same as in
Fig.~\ref{fig143677}
\label{fig040602}}
\end{figure*}

\clearpage
\begin{figure*}
\caption{{\bf ACO496J043333.17-131712.6} (G-17). The same as in
Fig.~\ref{fig143677}
\label{fig037145}}
\end{figure*}

\begin{figure*}
\caption{{\bf ACO496J043333.53-131852.6} (G-18). The same as in
Fig.~\ref{fig143677}
\label{fig033669}}
\end{figure*}

\clearpage
\begin{figure*}
\caption{{\bf ACO496J043334.54-131137.1} (G-19). The same as in
Fig.~\ref{fig143677}
\label{fig047746}}
\end{figure*}

\begin{figure*}
\caption{{\bf ACO496J043337.35-131520.2} (G-20). The same as in
Fig.~\ref{fig143677}
\label{fig041724}}
\end{figure*}

\clearpage
\begin{figure*}
\caption{{\bf ACO496J043338.22-131500.7} (G-21). The same as in
Fig.~\ref{fig143677}
\label{fig041729}}
\end{figure*}

\begin{figure*}
\caption{{\bf ACO496J043339.07-131319.7} (G-22). The same as in
Fig.~\ref{fig143677}
\label{fig044871}}
\end{figure*}

\clearpage
\begin{figure*}
\caption{{\bf ACO496J043339.72-131424.6} (G-23). The same as in
Fig.~\ref{fig143677}
\label{fig042581}}
\end{figure*}

\begin{figure*}
\caption{{\bf ACO496J043341.69-131551.8} (G-24). The same as in
Fig.~\ref{fig143677}
\label{fig039248}}
\end{figure*}

\clearpage
\begin{figure*}
\caption{{\bf ACO496J043342.10-131653.7} (G-25). The same as in
Fig.~\ref{fig143677}
\label{fig038089}}
\end{figure*}

\begin{figure*}
\caption{{\bf ACO496J043342.13-131258.8} (G-26). The same as in
Fig.~\ref{fig143677}
\label{fig043540}}
\end{figure*}

\clearpage
\begin{figure*}
\caption{{\bf ACO496J043342.83-130846.8} (G-27). The same as in
Fig.~\ref{fig143677}
\label{fig051603}}
\end{figure*}

\begin{figure*}
\caption{{\bf ACO496J043343.04-130514.1} (G-28). The same as in
Fig.~\ref{fig143677}
\label{fig058653}}
\end{figure*}

\clearpage
\begin{figure*}
\caption{{\bf ACO496J043343.04-125924.4} (G-29). The same as in
Fig.~\ref{fig143677}
\label{fig070277}}
\end{figure*}

\begin{figure*}
\caption{{\bf ACO496J043345.67-130542.2} (G-30). The same as in
Fig.~\ref{fig143677}
\label{fig058870}}
\end{figure*}

\clearpage
\begin{figure*}
\caption{{\bf ACO496J043346.71-131756.2} (G-31). The same as in
Fig.~\ref{fig143677}
\label{fig035258}}
\end{figure*}

\begin{figure*}
\caption{{\bf ACO496J043348.59-130558.3} (G-32). The same as in
Fig.~\ref{fig143677}
\label{fig058434}}
\end{figure*}

\clearpage
\begin{figure*}
\caption{{\bf ACO496J043349.08-130520.5} (G-33). The same as in
Fig.~\ref{fig143677}
\label{fig058478}}
\end{figure*}

\begin{figure*}
\caption{{\bf ACO496J043350.17-125945.4} (G-34). The same as in
Fig.~\ref{fig143677}
\label{fig070211}}
\end{figure*}

\clearpage
\begin{figure*}
\caption{{\bf ACO496J043351.54-131135.5} (G-35). The same as in
Fig.~\ref{fig143677}
\label{fig047971}}
\end{figure*}

\begin{figure*}
\caption{{\bf ACO496J043352.77-131523.8} (G-36). The same as in
Fig.~\ref{fig143677}
\label{fig040468}}
\end{figure*}

\clearpage
\begin{figure*}
\caption{{\bf ACO496J043355.55-131024.9} (G-37). The same as in
Fig.~\ref{fig143677}
\label{fig049704}}
\end{figure*}

\begin{figure*}
\caption{{\bf ACO496J043356.18-125913.1} (G-38). The same as in
Fig.~\ref{fig143677}
\label{fig070215}}
\end{figure*}

\clearpage
\begin{figure*}
\caption{{\bf ACO496J043359.03-130626.7} (G-39). The same as in
Fig.~\ref{fig143677}
\label{fig056580}}
\end{figure*}

\begin{figure*}
\caption{{\bf ACO496J043401.57-131359.7} (G-40). The same as in
Fig.~\ref{fig143677}
\label{fig042823}}
\end{figure*}

\clearpage
\begin{figure*}
\caption{{\bf ACO496J043403.19-131310.6} (G-41). The same as in
Fig.~\ref{fig143677}
\label{fig043592}}
\end{figure*}

\begin{figure*}
\caption{{\bf ACO496J043408.50-131152.7} (G-42). The same as in
Fig.~\ref{fig143677}
\label{fig046374}}
\end{figure*}

\clearpage
\begin{figure*}
\caption{{\bf ACO496J043410.60-130756.7} (G-43). The same as in
Fig.~\ref{fig143677}
\label{fig053384}}
\end{figure*}

\begin{figure*}
\caption{{\bf ACO496J043413.00-131003.5} (G-44). The same as in
Fig.~\ref{fig143677}
\label{fig049571}}
\end{figure*}

\clearpage
\begin{figure*}
\caption{{\bf ACO496J043413.08-131231.6} (G-45). The same as in
Fig.~\ref{fig143677}
\label{fig045854}}
\end{figure*}

\begin{figure*}
\caption{{\bf ACO496J043415.37-130823.5} (G-46). The same as in
Fig.~\ref{fig143677}
\label{fig052605}}
\end{figure*}

\clearpage
\begin{figure*}
\caption{{\bf ACO496J043411.72-131130.2} (G-A1). The same as 
panels (a) -- (e) in Fig.~\ref{fig143677}
\label{fig048282}}
\end{figure*}

\begin{figure*}
\caption{{\bf ACO496J043414.54-131303.0} (G-A2). The same as 
panels (a) -- (e) in Fig.~\ref{fig143677}
\label{fig045383}}
\end{figure*}

\section{Stability and biases of the spectral fitting technique}

In order to assess the reliability and precision of the stellar
population parameters found by the pixel fitting procedure under
different circumstances, we have conducted a number of tests using
Monte-Carlo simulations and published datasets. 

\subsection{Multiplicative polynomial continuum}
The fitting procedure includes a multiplicative function whose role is to
absorb the flux calibration errors, both in the population models and in the
observations, and the effect of the extinction (internal and Galactic).
Whenever the flux calibration can be trusted, this function can be used to
determine the extinction. Since the FLAMES/Giraffe data are not flux
calibrated, the usage of a multiplicative continuum is absolutely
required. This feature makes the method comparable to spectrophotometric
indices, which are also insensitive to the shape of the continuum (they are
similar to equivalent widths).

The degree of the polynomial is chosen to remove the signatures of
continuum mismatch at the scale of 60~\AA\ and above. The effect of the
multiplicative function is thoroughly discussed in Koleva et al. (2008).
There is no systematic bias of the results (age, metallicity, and kinematics)
as a function of the polynomial degree, but when the degree
increases, the well of the $\chi^2$ becomes deeper and sharper, allowing a
better identification of the absolute minimum and giving more precise
measurements. There is no degeneracy between the coefficients of the
polynomial and the other parameters, even between the age and the low-degree
coefficients. 

\begin{figure}
\includegraphics[width=8cm]{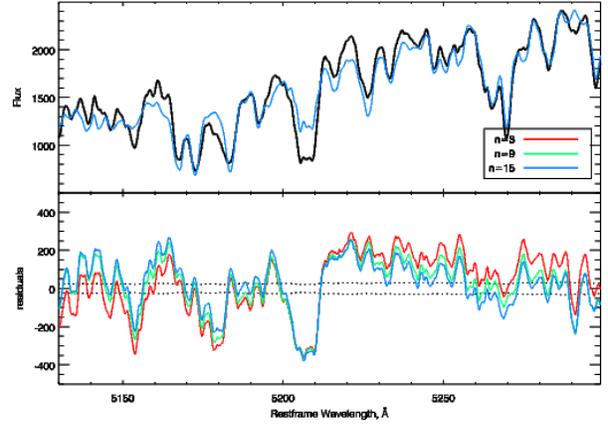}
\caption{Residuals of the fitting for a late-type star, an object with 
a strong MgH (5208\AA) molecular band. The top panel presents the rest-frame
spectrum of the star (black) and its best fitting PEGASE.HR template for
$n=15$ (blue), the bottom panel displays the fitting residuals for the three
different multiplicative polynomial continua ($n=3, 9, 15$). The visual
scale of the flux axis for the residuals is doubled compared to the top
panel. The 1$\sigma$ flux uncertainty is shown as black dotted lines. Both
spectrum and residuals are smoothed using a boxcar window of 15 pixels.
\label{figmgh}}
\end{figure}

Koleva et al. (2008) also stress that unlike the ``rectification'' or
filtering approach of, e.g., Mathis et al. (2006) or Wolf et al. (2007) the
information contained in the broad features like MgH, TiO, or CN bands is not
altered by the continuum. For the FLAMES/Giraffe data this is demonstrated in
Fig.~\ref{figmgh}, where the spectrum of a late-type star with a deep MgH
5208\AA\ molecular band is shown with the residuals for its best-fitting
PEGASE.HR stellar populations for the three different levels of the
multiplicative polynomial continuum $n=$3, 9, and 15. The stellar spectrum
has been obtained in the frame of another observational programme using the
same FLAMES/Giraffe setup as in our study. This fitting is not physically
correct because an individual stellar spectrum cannot be represented by the
spectrum of a stellar population, however, this test demonstrates very well
that the fitting residuals of the MgH band do not change at all for very
strong variations of the multiplicative continuum order.

We have followed the recipe described in Sect.~A2.3 of Chilingarian et al.
(2007a) to determine the minimal order of the multiplicative polynomial
continuum sufficient for fitting FLAMES/Giraffe data.

\begin{figure}
\includegraphics[width=8cm]{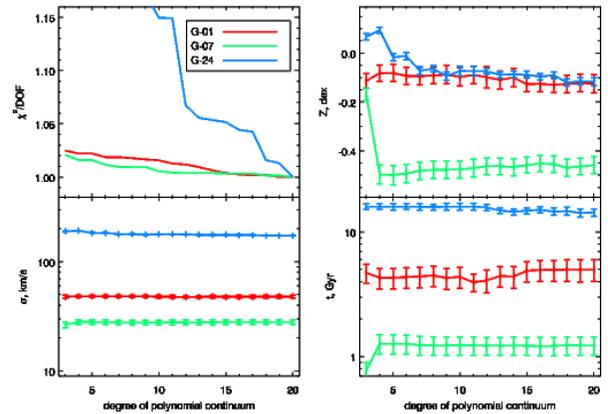}
\caption{Stability of the estimations of the velocity dispersion (bottom left), 
metallicity (top right), and age (bottom right) with respect to the
multiplicative polynomial continuum order ($n$) for three galaxies with
different stellar populations. $\chi^2$ (top left) is normalised by the
value, corresponding to $n = 20$ for every object.
\label{figstabnp}}
\end{figure}

We have taken three objects from our sample having young (G-01),
intermediate (G-07), and old (G-24) ages and fit their spectra varying the
polynomial continuum order from $n=3$ to $n=20$. Apart from an old stellar
population, G-24 also exhibits a significantly super-solar [Mg/Fe] abundance
ratio, providing the possibility to test how a multiplicative
polynomial continuum may reduce the template mismatch. Figure~\ref{figstabnp}
demonstrates the relative changes of the $\chi^2$ as well as trends of the
kinematical and stellar population parameters.

For $n \geq 15$, none of the parameters shows significant
changes. For the two galaxies (G-01 and G-07) having nearly solar [Mg/Fe]
abundance ratios, the trends of age, metallicity, velocity dispersion, and
$\chi^2$ estimates are negligible for $n>3$, however, for the
third galaxy, the behaviour of the parameters becomes stable only for $n>8$,
although $\chi^2$ exhibits a sharp drop at a 10~percent level at $n=12$. At
the same time, the reduced $\chi^2$ remains almost twice its value for G-01
or G-07.

\begin{figure}
\includegraphics[width=8cm]{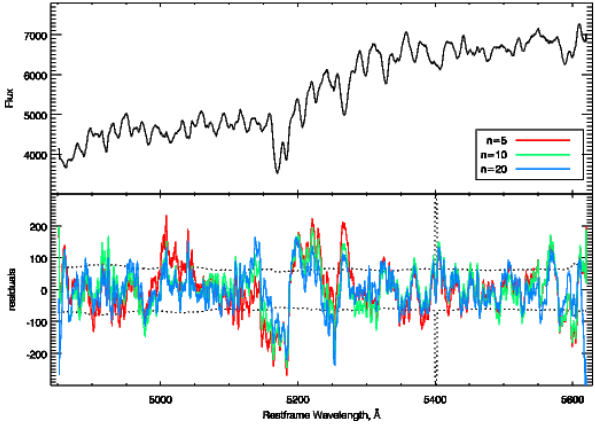}
\caption{Residuals of the fitting for G-24, an object with highly 
overabundant [$\alpha$/Fe]. The top panel presents the rest-frame spectrum
of the galaxy. The bottom panel displays the fitting residuals for the three
different multiplicative polynomial continua ($n=5, 10, 20$). The visual scale
of the flux axis for the residuals is increased by a factor 8 compared to the
top panel. The 1$\sigma$ flux uncertainty is shown as black dotted lines.
Both spectrum and residuals are smoothed using a boxcar window of 15 pixels.
\label{residG24}}
\end{figure}

The analysis of the fitting residuals for G-24 shows that increasing the
continuum order allows us to reduce the template mismatch for spectral regions
containing numerous blends of iron or $\alpha$-element lines. In
Figure~\ref{residG24}, we present the fitting residuals for G-24 for the three
different orders of the multiplicative polynomial continuum: 5, 10, and 20.
One notices that the iron-dominated region between 5005 and 5050~\AA\ is seen
very well in the residuals for $n=5$, but almost disappears for $n=10$. Changes
of the residuals become insignificant for $n > 14$, therefore, we have chosen
$n=15$ as the optimal multiplicative polynomial continuum order when processing
FLAMES/Giraffe data in the LR4 setup.

\subsection{Non-solar [Mg/Fe] abundance ratios}

Quite a high fraction of massive objects in our sample exhibit
supersolar values of [Mg/Fe], therefore, we pose the following principal
questions for the validation of our results. Does our technique produce
biased estimations of single stellar population (hereafter SSP), and
equivalent ages and metallicities in the cases of non-solar [Mg/Fe]
abundance ratios? Do our results depend on the presence of H$\beta$?  If
there are biases, is it still possible to apply some empirical
corrections?

Presently, there are no publicly available models of spectral energy
distributions of synthetic stellar populations for non-solar
$\alpha$-element abundance ratios. Therefore we use published spectral
data, where the age and metallicity can be estimated using both Lick
indices and pixel fitting.  We have selected in the SDSS DR6
(Adelman-McCarthy et al. 2008) 1200 spectra of galaxies having redshifts
$z<0.033$, colours $(g' - r')_{\mbox{fib}} > 0.9$, and signal-to-noise
ratios $S/N > 20$. The spectral resolution information (Gaussian width
of the instrumental LSF at every wavelength) is available for each
individual spectrum. For $\lambda <5900$\AA, it increases smoothly from
blue to red with values around $\sigma_{\mbox{inst}} \approx
75$~km~s$^{-1}$.

We have not applied any selection criteria based on galaxy morphology.
Most of the galaxies (about 95\%) exhibit at least faint emission
lines, the strongest one usually being [OIII] ($\lambda = 5007$)~\AA. If the
equivalent width of the [OIII] line was reported to be greater than zero in the SDSS
data, we substituted the observed fluxes with those of the
best-fitting PEGASE.HR templates in the 8\AA -wide regions around
$H\delta$, $H\gamma$, [OIII]$\lambda 4363$, [NI]$\lambda 5199$, and
10\AA -wide regions around $H\beta$, [OIII]$\lambda\lambda 4959-5007$. This
was done to avoid contamination of the measurements of Lick indices
by emission lines. We decided to use this approach rather than applying
corrections based on the intensities of other well-measured emission
lines multiplied by empirical coefficients because the latter technique may
lead to spurious results due to: (a) extinction inside the object, affecting
Balmer decrement, and, therefore, the H$\alpha$/H$\beta$ ratio; or (b)
possible activity in the galactic nucleus, affecting the [OIII]/H$\beta$
ratio, which is unaffected by the effects of extinction. Although replacing
up-to 30 percent of the counts in the index band by the best-fitting may
slightly bias the measurements of the H$\beta$ and Mg$b$ indices toward
best-fitting PEGASE.HR models, all possible biases correlated or
anticorrelated with the values of the [Mg/Fe] abundance ratio should be clearly
revealed at least on a qualitative level.

To measure Lick indices we have degraded the spectral resolution of the
SDSS by convolving the original spectra with a Gaussian of width
$\sigma_{degr} = \sqrt{\sigma_{Lick}^2 - \sigma_{SDSS}^2 -
\sigma_{g}^2}$, where $\sigma_{g}$ is velocity dispersion of the galaxy,
$\sigma_{SDSS}$ is the width of the SDSS LSF and $\sigma_{Lick}$ is the
resolution needed to measure Lick indices, both depend on the index
considered. If the value under the square root turned out to be
negative, we did no degradation. Instead, the $\sigma$-correction
according to Kuntshner (2004) was applied to the corresponding
measurements of Lick indices using $\sigma_{corr} =
\sqrt{\sigma_{SDSS}^2 + \sigma_{g}^2 -\sigma_{Lick}^2}$.  The quality of
the flux calibration of the SDSS DR6 spectra (Adelman-McCarthy et
al. 2008; Schlegel 2007, private communication) has been dramatically
improved since SDSS DR5, reaching a precision of better than one percent
for relative fluxes, therefore, we consider that the absolute values of
Lick indices can be trusted and not only the relative trends, as e.g.,
in Clemens et al. (2006), based on SDSS DR3.

The values of ages, metallicities, and [Mg/Fe] abundance ratios have
been obtained by inverting the grids of SSP models for H$\beta$, Mg$b$
and $\langle$Fe$\rangle$ Lick indices by Thomas et al. (2003). We have
first inverted the $\langle$MgFe$\rangle$ - H$\beta$ grid to derive the
age and mean metallicity; then another inversion of the Mg$b$ -
$\langle$Fe$\rangle$ grid for the age value found in the previous step
has been done to obtain [Mg/Fe] abundance ratios.  In the final sample
we have kept 848 SDSS spectra, having 1.4$<$H$\beta <$2.6~\AA\ and 2.0$<
\langle$MgFe$\rangle <$4.2~\AA, corresponding to intermediate-age and
old stellar populations.

The spectral fitting has been performed in the three following
wavelength ranges (restframe): (1) 4100 -- 5800~\AA, (2) 4800 --
5600~\AA, (3) 4880 -- 5680~\AA. We chose the first interval to
cover the broad wavelength range often used in the intermediate
resolution ground-based spectroscopic observations aimed at studies of
stellar kinematics and stellar populations (GMOS at Gemini, MPFS at
the Russian 6-m telescope, etc.). The second and third ones correspond
to the wavelength ranges of the LR4 setup of FLAMES/Giraffe for
objects at the redshift of \object{Abell~496}, including and excluding the age
sensitive H$\beta$ absorption feature.

The radial velocities provided in the SDSS and found by the spectral
fitting procedure coincide within the error-bars, rarely exceeding a
few km~s$^{-1}$. The velocity dispersions (provided by the SDSS for
824 of the 1200 galaxies) remain consistent within uncertainties for
$\sigma < 170$~km~s$^{-1}$, but for larger values of $\sigma$ there is
a prominent trend: the SDSS values turn out to be lower than our
measurements, reaching $\Delta \sigma \sim$20~km~s$^{-1}$ for $\sigma
\sim 300$~km~s$^{-1}$.  This effect can be explained by the $\sigma$ -
metallicity degeneracy mentioned in Chilingarian (2006) and
Chilingarian et al. (2007a) -- galaxies with large velocity dispersions
tend to be more metal-rich, but the quality of templates used by the
SDSS team to measure velocity dispersions was not as good as
PEGASE.HR, leading to underestimates of velocity dispersions for high
metallicities. A correlation between $\Delta \sigma$ and metallicity
confirms this hypothesis.

\begin{figure} 
\includegraphics[width=8cm]{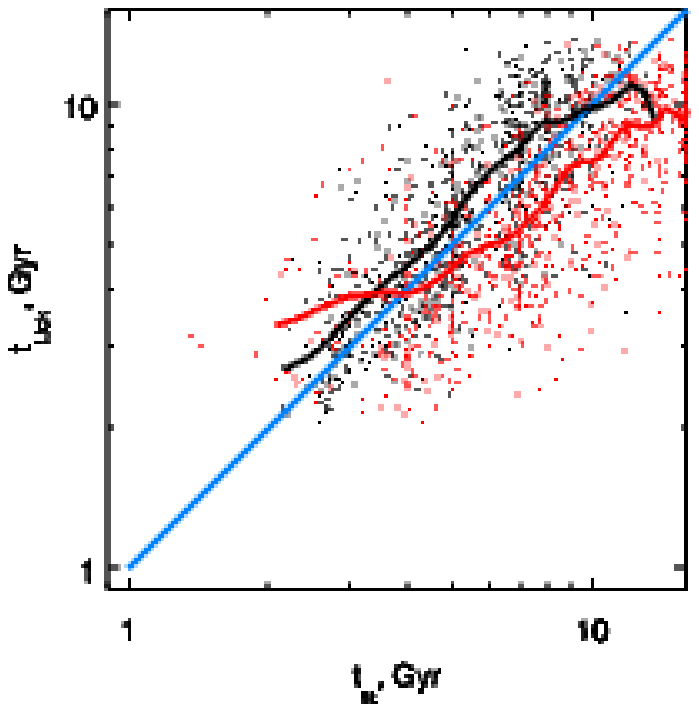} \\
\includegraphics[width=8cm]{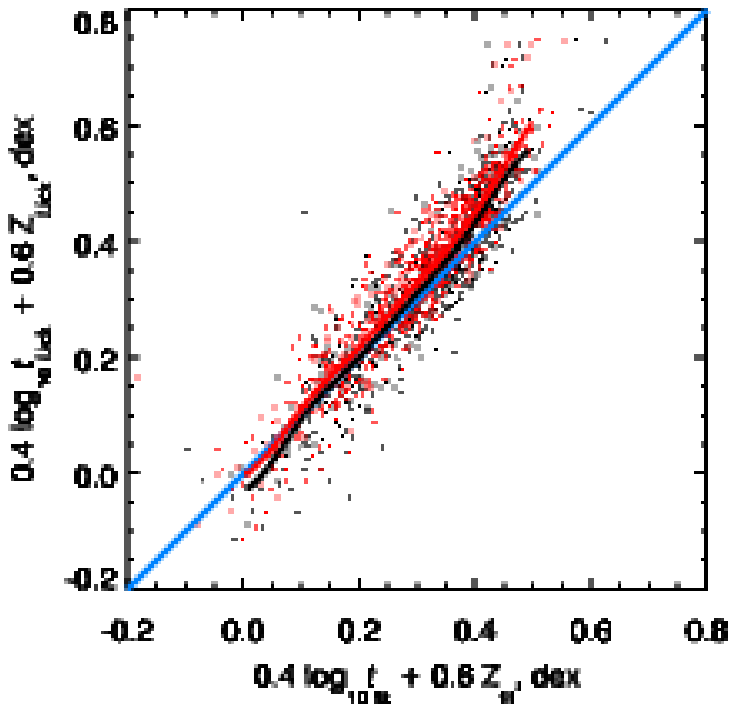}
\caption{Comparison of SSP ages (top panel) and combinations of age
and metallicity, stable against the age-metallicity degeneracy (bottom
panel), obtained using Lick indices and spectral fitting. Black and
red points correspond to the spectral fitting in the two following
wavelength ranges respectively: 4100 -- 5800~\AA\ and 4800 --
5600~\AA.
\label{figageLickFit}}
\end{figure}

The top panel of Fig.~\ref{figageLickFit} shows the comparison between
the age determinations based on the Lick indices and those derived
from the stellar population fitting technique. The agreement between
the measurements is relatively good. The spread of points has two main
sources: (1) statistical uncertainties on the measurements of the Lick
indices and on the results of the spectral fitting; and (2) age~-
metallicity degeneracy. In order to demonstrate the agreement between
both methods for stellar population parameter determination, we plot a
combination of age and metallicity, $0.4 \log_{10} \mbox{t} + 0.6
\mbox{Z}$ (according to Worthey et al. 1994) on the bottom panel of
Fig.~\ref{figageLickFit}. A particularly good agreement is observed
between the results obtained from the spectral fitting in the two
wavelength ranges and from the inversion of grids of Lick indices
(H$\beta$ and $\langle$MgFe$\rangle$).

\begin{figure} 
\includegraphics[width=8cm]{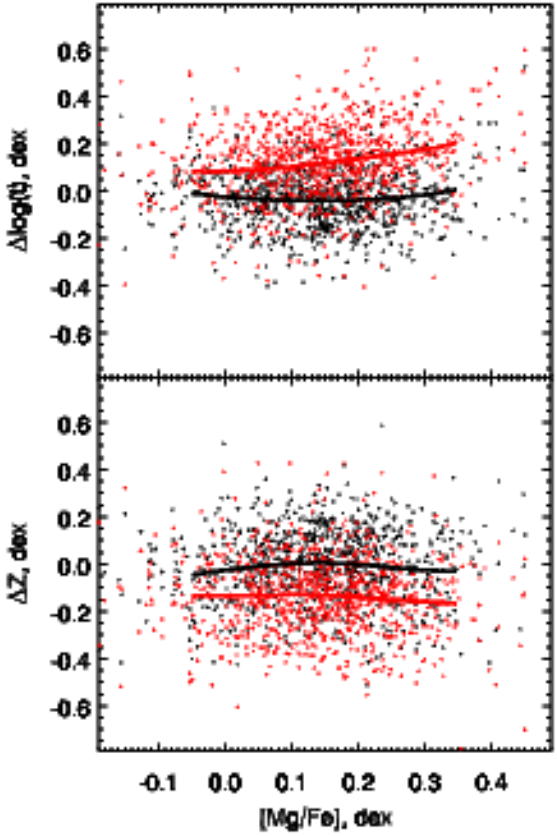}
\caption{Difference of age (top panel) and metallicity (bottom panel)
determinations from the spectral fitting and Lick indices against 
[Mg/Fe] abundance ratios. Black and red colours are the same as in 
Fig.~\ref{figageLickFit}.
\label{figSDSSagemetMgFe}}
\end{figure}

A particularly important result of this study is the fact that
the age and metallicity determinations obtained from the spectral
fitting are independent from the non-solar [Mg/Fe] abundance ratios of
the stellar populations. This is illustrated in
Fig.~\ref{figSDSSagemetMgFe}. The ordinate axes represent differences
(``fit'' - ``Lick'') between the decimal logarithms of age (top panel)
and metallicity (bottom panel) measurements for objects with various
abundance ratios of [Mg/Fe]. One can notice the total absence of
correlation with [Mg/Fe] ratio for both age and metallicity
measurements, although there are slight systematic shifts
($\sim$0.1~dex) due to age-metallicity degeneracy effects.  A similar
result has been found for the spectra of star clusters by Koleva et
al. (2007).  Since the black line indicating the mean difference for
the large wavelength range is very near the zero value, we can ensure
that the measurements of the Lick indices do not suffer from
systematic offsets.
Results obtained from the spectral fitting in the 4800 -- 5600~\AA\ and 
4880 -- 5680~\AA\ wavelength ranges are absolutely consistent, showing 
systematic differences neither in age, nor in metallicity estimates.

We have performed the scanning of the age-metallicity parameter space to
locate the absolute minima of $\chi^2$ for 600 SDSS spectra by fixing
age and metallicity values and fitting only the kinematics and
multiplicative polynomial continuum (as explained in Sects.~3 and
B.1). All three wavelength ranges have been processed. We did this to
verify how the spectral fitting procedure based on a non-linear
minimisation works in the case of complex shapes of the minima. The
agreement between the results of the fitting and scanning is very good
for both age and metallicity. In 99 percent of the cases the
``best-scanning'' values coincide within the size of the bin in the
$(t,Z)$ space with the ``best-fitting'' values. In the remaining
1~percent, the determined ages are as old as the oldest population in
the grid of models, so matching between the values cannot be fully
trusted.

The main conclusion we draw from this Appendix is that {\sl our
spectral fitting procedure is quite robust and produces unbiased age
and metallicity estimates even for strongly non-solar [Mg/Fe]
abundance ratios, although it is based on models having
[Mg/Fe]=0}. This allows us to exploit the proposed technique to study the
kinematics and stellar populations not only of dwarf galaxies, known
to exhibit solar [Mg/Fe] element ratios, but also of intermediate
luminosity and giant early-type galaxies as well as bulges of spirals,
known to be overabundant in $\alpha$-elements.

\section{Redshifts of field galaxies}
In Table~\ref{tabfldz} we present the coordinates and redshifts of the 54 field
galaxies in the direction of the \object{Abell~496} cluster, observed with 
FLAMES/Giraffe. We do not discuss these results here. We can
mention a
large concentration of bright galaxies with redshifts around 0.17 and 0.19,
probably members of an unknown cluster, located nearly on the same line of
sight as \object{Abell~496}, and already detected by Durret et al. (2000) as structure
\#9 (in their Table~1).

\begin{table}
\caption{Redshifts of the field galaxies in the direction of \object{Abell~496}.
Presence of emission lines is indicated in the last column of the table.
\label{tabfldz}}
\begin{tabular}{llcc}
\hline
\hline
N & IAU Name & z & em.l\\
\hline
01 & ACO496J043254.82-130920.0 & 0.0743 & + \\
02 & ACO496J043257.55-130923.2 & 0.3164 & + \\
03 & ACO496J043259.43-131035.9 & 0.1672 & + \\
04 & ACO496J043300.38-131043.3 & 0.0744 & + \\
05 & ACO496J043300.91-130944.6 & 0.2219 & + \\
06 & ACO496J043301.05-130428.3 & 0.3451 & + \\
07 & ACO496J043303.28-130659.2 & 0.2631 & + \\
08 & ACO496J043306.56-130404.7 & 0.3330 & + \\
09 & ACO496J043309.11-131434.0 & 0.0993 & + \\
10 & ACO496J043310.78-131532.9 & 0.1734 & + \\
11 & ACO496J043315.04-130112.4 & 0.2212 & + \\
12 & ACO496J043315.64-131230.6 & 0.2240 & + \\
13 & ACO496J043317.87-130450.2 & 0.4535 & + \\
14 & ACO496J043317.97-131626.5 & 0.1275 & + \\
15 & ACO496J043319.29-131227.7 & 0.2802 & + \\
16 & ACO496J043319.90-130900.8 & 0.1744 & - \\
17 & ACO496J043321.70-130804.8 & 0.3147 & + \\
18 & ACO496J043323.02-130633.7 & 0.4551 & + \\
19 & ACO496J043324.34-130522.1 & 0.2546 & + \\
20 & ACO496J043327.88-130800.6 & 0.2735 & + \\
21 & ACO496J043328.44-131858.1 & 0.3588 & + \\
22 & ACO496J043328.51-130925.6 & 0.2790 & + \\
23 & ACO496J043331.18-130435.3 & 0.2630 & + \\
24 & ACO496J043332.92-130659.5 & 0.0844 & + \\
25 & ACO496J043337.17-125808.5 & 0.1336 & + \\
26 & ACO496J043337.21-130445.1 & 0.0842 & + \\
27 & ACO496J043337.87-131841.6 & 0.1958 & - \\
28 & ACO496J043338.45-130239.4 & 0.0559 & + \\
29 & ACO496J043340.03-131342.2 & 0.3582 & + \\
30 & ACO496J043340.63-131925.0 & 0.2730 & + \\
31 & ACO496J043342.84-131230.9 & 0.3097 & + \\
32 & ACO496J043346.79-131310.4 & 0.3577 & + \\
33 & ACO496J043346.96-130948.0 & 0.1967 & - \\
34 & ACO496J043348.07-130231.0 & 0.1795 & - \\
35 & ACO496J043348.09-131509.4 & 0.2732 & + \\
36 & ACO496J043348.96-130259.5 & 0.4946 & + \\
37 & ACO496J043350.42-130214.7 & 0.1790 & + \\
38 & ACO496J043354.89-131734.4 & 0.3451 & + \\
39 & ACO496J043356.50-130121.2 & 0.0851 & + \\
40 & ACO496J043356.68-131346.2 & 0.1918 & + \\
41 & ACO496J043357.22-130609.7 & 0.1798 & + \\
42 & ACO496J043357.62-130419.5 & 0.0695 & + \\
43 & ACO496J043358.32-131428.7 & 0.1606 & + \\
44 & ACO496J043358.98-130301.5 & 0.2589 & + \\
45 & ACO496J043401.51-131544.2 & 0.1795 & - \\
46 & ACO496J043402.18-130637.7 & 0.1797 & - \\
47 & ACO496J043403.80-130149.4 & 0.1832 & + \\
48 & ACO496J043404.73-130203.5 & 0.1894 & + \\
49 & ACO496J043405.98-130722.7 & 0.1912 & + \\
50 & ACO496J043408.53-131257.8 & 0.1714 & + \\
51 & ACO496J043409.64-130742.6 & 0.5077 & + \\
52 & ACO496J043410.28-130841.9 & 0.1909 & + \\
53 & ACO496J043411.96-130522.4 & 0.1921 & + \\
54 & ACO496J043416.79-130543.1 & 0.4347 & + \\
\hline
\hline
\end{tabular}
\end{table}

\end{document}